\documentclass{aa}

\usepackage{graphicx}
\usepackage{txfonts}
\usepackage{natbib}     
\usepackage{xcolor} 
\usepackage{threeparttable} 
\usepackage{float}
\usepackage{threeparttable} 
\usepackage{placeins} 

\usepackage{caption}
\usepackage{subcaption}
\usepackage{float}

\bibpunct{(}{)}{;}{a}{}{,} 

\begin{document}

\title{Atmospheric characterization of the ultra-hot Jupiter WASP-33b}
\subtitle{Detection of Ti and V emission lines and retrieval of a broadened line profile}

\author{D.~Cont\inst{1}
        \and
        F.~Yan\inst{2,1}
        \and
        A.~Reiners\inst{1}
        \and
        L.~Nortmann\inst{1}
        \and
        K.~Molaverdikhani\inst{3,4,5,6}
        \and
        E.~Pall\'e\inst{7,8}
        \and
        Th.~Henning\inst{6}    
        \and
        I.~Ribas\inst{9,10}
        \and
        A.~Quirrenbach\inst{5}
        \and
        J.~A.~Caballero\inst{11}
        \and
        P.~J.~Amado\inst{12}
        \and
        S.~Czesla\inst{13}
        \and
        F.~Lesjak\inst{1}
        \and
        M.~L\'opez-Puertas\inst{12}
        \and
        P.~Molli\`ere\inst{6}
        \and
        D.~Montes\inst{14}
        \and
        G.~Morello\inst{7,8}
        \and
        E.~Nagel\inst{13,15}
        \and 
        S.~Pedraz\inst{16} 
        \and
        A.~S\'anchez-L\'opez\inst{17}
        \\
}

\institute{Institut f\"ur Astrophysik und Geophysik, Georg-August-Universit\"at, Friedrich-Hund-Platz 1, 37077 G\"ottingen, Germany\\
        \email{david.cont@uni-goettingen.de, yanfei@ustc.edu.cn}
        \and
        Department of Astronomy, University of Science and Technology of China, Hefei 230026, China
        \and
        Universit\"ats-Sternwarte, Ludwig-Maximilians-Universit\"at M\"unchen, Scheinerstrasse 1, 81679 M\"unchen, Germany
        \and
        Exzellenzcluster Origins, Boltzmannstraße 2, 85748 Garching, Germany
        \and
        Landessternwarte, Zentrum f\"ur Astronomie der Universit\"at Heidelberg, K\"onigstuhl 12, 69117 Heidelberg, Germany      
        \and
        Max-Planck-Institut f{\"u}r Astronomie, K{\"o}nigstuhl 17, 69117 Heidelberg, Germany     
        \and
        Instituto de Astrof{\'i}sica de Canarias (IAC), Calle V{\'i}a Lactea s/n, 38200 La Laguna, Tenerife, Spain
        \and
        Departamento de Astrof{\'i}sica, Universidad de La Laguna, 38026  La Laguna, Tenerife, Spain
        \and
        Institut de Ci\`encies de l'Espai (CSIC-IEEC), Campus UAB, c/ de Can Magrans s/n, 08193 Bellaterra, Barcelona, Spain
        \and
        Institut d'Estudis Espacials de Catalunya (IEEC), 08034 Barcelona, Spain
        \and
        Centro de Astrobiolog{\'i}a (CAB), CSIC-INTA, Camino bajo del castillo s/n, Campus ESAC, 28692 Villanueva de la Ca{\~n}ada, Madrid, Spain
        \and
        Instituto de Astrof{\'i}sica de Andaluc{\'i}a (IAA-CSIC), Glorieta de la Astronom{\'i}a s/n, 18008 Granada, Spain
        \and
        Th{\"u}ringer Landessternwarte Tautenburg, Sternwarte 5, 07778 Tautenburg, Germany
        \and
Departamento de F\'{i}sica de la Tierra y Astrof\'{i}sica 
        and IPARCOS-UCM (Instituto de F\'{i}sica de Part\'{i}culas y del Cosmos de la UCM), 
        Facultad de Ciencias F\'{i}sicas, Universidad Complutense de Madrid, 28040, Madrid, Spain
        \and
        Hamburger Sternwarte, Universit{\"a}t Hamburg, Gojenbergsweg 112, 21029 Hamburg, Germany
        \and
        Centro Astron{\'o}mico Hispano en Andaluc{\'i}a, Observatorio de Calar Alto, Sierra de los Filabres, 04550 G{\'e}rgal, Spain
        \and
        Leiden Observatory, Universiteit Leiden, Postbus 9513, 2300 RA, Leiden, The Netherlands
}

\date{Received 15 June 2022 / Accepted 17 September 2022}


\abstract
{Ultra-hot Jupiters are highly irradiated gas giant exoplanets on close-in orbits around their host stars. The dayside atmospheres of these objects strongly emit thermal radiation due to their elevated temperatures, making them prime targets for characterization by emission spectroscopy. We analyzed high-resolution spectra from CARMENES, HARPS-N, and ESPaDOnS taken over eight observation nights to study the emission spectrum of WASP-33b and draw conclusions about its atmosphere. By applying the cross-correlation technique, we detected the spectral signatures of \ion{Ti}{i}, \ion{V}{i}, and a tentative signal of \ion{Ti}{ii} for the first time via emission spectroscopy. These detections are an important finding because of the fundamental role of Ti- and V-bearing species in the planetary energy balance. Moreover, we assessed and confirm the presence of OH, \ion{Fe}{i}, and \ion{Si}{i} from previous studies. The spectral lines are all detected in emission, which unambiguously proves the presence of an inverted temperature profile in the planetary atmosphere. By performing retrievals on the emission lines of all the detected species, we determined a relatively weak atmospheric thermal inversion extending from approximately 3400\,K to 4000\,K. We infer a supersolar metallicity close to 1.5\,dex in the planetary atmosphere, and find that its emission signature undergoes significant line broadening with a Gaussian FWHM of about 4.5\,km\,s$^{-1}$. Also, we find that the atmospheric temperature profile retrieved at orbital phases far from the secondary eclipse is about 300\,K to 700\,K cooler than that measured close to the secondary eclipse, which is consistent with different day- and nightside temperatures. Moreover, retrievals performed on the emission lines of the individual chemical species lead to consistent results, which gives additional confidence to our retrieval method. Increasing the number of species included in the retrieval and expanding the set of retrieved atmospheric parameters will further advance our understanding of exoplanet atmospheres.}

\maketitle

%

\section{Introduction}

In-depth characterization of exoplanet atmospheres is an emerging field in astronomy. Most known exoplanets have no analogs in our Solar System and therefore challenge our understanding of their formation and evolution. In recent years, observations and theoretical work have increased our knowledge about the subclass of ultra-hot Jupiters (UHJs; \citealt{Parmentier2018}), which are gas giant planets with the highest equilibrium temperatures identified to date ($T_\mathrm{eq}$\,$\ge$\,2200\,K). Located on close-in orbits, they are expected to be tidally locked to their host stars with highly irradiated daysides. The extreme irradiation regime in combination with a synchronous rotation causes a strong temperature contrast between the permanent day- and nightsides. Therefore, the atmospheric composition is significantly different between the two planetary hemispheres \citep[e.g.,][]{Arcangeli2018, BellCowan2018, KomacekTan2018, TanKomacek2019, Molaverdikhani2020}. UHJ daysides are likely dominated by atomic species with a high degree of ionization, while most of the molecules are predicted to dissociate \citep{Kitzmann2018, Lothringer2018}. In contrast, the planetary nightsides harbor a large variety of molecules and can be covered by clouds with complex molecular chemistry \citep{Helling2019, Gao2020, GaoPoweell2021}.
        
The atmospheric composition of UHJs has been extensively studied in recent years. For instance, neutral or ionic species of H, Li, O, Na, Mg, Si, K, Ca, Sc, Ti, V, Cr, Mn, Fe, Sr, and Y have been identified in the atmospheres of several UHJs, including KELT-9b \citep{Yan&Henning2018, Hoeijmakers2018, Hoeijmakers2019, Wyttenbach2020, Borsa2021}, WASP-33b \citep{Yan2019, Yan2021, Nugroho2020_Fe, Borsa2021_1, Cont2021, Cont2022, Herman2022}, WASP-189b \citep{Yan2020, Stangret2022, Prinoth2022}, WASP-76b \citep{Seidel2019, Ehrenreich2020, Tabernero2021, Deibert2021, CasasayasBarris2021, Kesseli2022}, and KELT-20b/MASCARA-2b \citep{Casasayas-Barris2018, Casasayas-Barris2019, Stangret2020, Nugroho2020_KELT20b, Hoeijmakers2020, Rainer2021, Yan2022, Borsa2022}. Also, the spectral signatures of an increasing number of molecular species such as CO, HCN, H$_2$O, OH, SiO, or TiO have been identified in UHJ atmospheres \citep[e.g.,][]{Nugroho2021, Cont2021, Landman2021, Fu2022, Prinoth2022, SanchezLopez2022, Yan2022_CO, vanSluijs2022, Lothringer2022}. Detecting the spectral lines of different species not only allows for the characterization of atmospheric chemistry, but also provides the opportunity to constrain additional properties of UHJs. For example, strong day-to-night winds \citep[e.g.,][]{Louden2015, Brogi2016, Alonso-Floriano2019}, atmospheric mass loss \citep[e.g.,][]{Yan&Henning2018, Wyttenbach2020}, or rain-out from the gas phase on the planetary nightside \citep{Ehrenreich2020} have been identified. Moreover, the composition of exoplanet atmospheres can provide insights into the planetary formation and migration history, especially if the concentrations of the most important carbon- and oxygen-bearing species are well determined \citep{Mordasini2016, Madhusudhan2019}.

        \begin{table}[h!]
                \caption{Parameters of the \object{WASP-33} system.}             
                \label{tab-parameters}                           
                \centering                                       
                \renewcommand{\arraystretch}{1.3} 
                \begin{threeparttable}
                        \begin{tabular}{l l l}                       
                                \noalign{\smallskip}
                                \hline\hline                             
                                \noalign{\smallskip}
                                Parameter & Symbol (Unit) & Value \\     
                                \noalign{\smallskip}
                                \hline                                   
                                \noalign{\smallskip}
                                \textit{Planet} & &  \\ 
                                \noalign{\smallskip}
                                Radius  \tablefootmark{a}               & $R_\mathrm{p}$ ($R_\mathrm{Jup}$) & $1.679_{-0.030}^{+0.019}$ \\
                                Orbital period \tablefootmark{b}      & $P_\mathrm{orb}$ (d) & 1.219870897 \\
                                Transit epoch (BJD) \tablefootmark{b}  & $T_\mathrm{0}$ (d) & 2454163.22449\\
                                Systemic velocity \tablefootmark{c}    & $\varv_\mathrm{sys}$ (km\,s$^{-1}$) & $-3.0\pm0.4$\\
                                RV semi-amplitude \tablefootmark{a} & $K_\mathrm{p}$ (km\,s$^{-1}$) & $231\pm3$\\
                                Ingress duration \tablefootmark{d}     & $T_\mathrm{ingress}$ (h) & $0.298\pm0.005$\\
                                Transit duration \tablefootmark{d}     & $T_\mathrm{transit}$ (h) & $2.743\pm0.005$\\
                                Surface gravity \tablefootmark{d}     & log $g$ (cgs) & 3.46\\
                                \noalign{\smallskip} \hline \noalign{\smallskip}
                                \textit{Star} & &  \\  
                                \noalign{\smallskip}
                                Radius \tablefootmark{a} & $R_*$ ($R_\mathrm{\sun}$) & $1.509_{-0.027}^{+0.016}$\\ 
                                Effective temperature \tablefootmark{e} & $T_\mathrm{eff}$ (K) & $7430\pm100$\\                 
                                Rotational velocity \tablefootmark{f} & $\varv_\mathrm{rot}\sin{i_*}$ &  $86.63_{-0.32}^{+0.37}$\\ 
                                & (km\,s$^{-1}$) & \\
                                \noalign{\smallskip}
                                \hline                                   
                        \end{tabular}
                        \tablefoot{
                                \tablefoottext{a}{\cite{Lehmann2015} with parameters from \cite{Kovacs2013}}, \tablefoottext{b}{\cite{Maciejewski2018}},
                                \tablefoottext{c}{\cite{Nugroho2017}},
                                \tablefoottext{d}{\cite{Kovacs2013}},
                                \tablefoottext{e}{\cite{Collier-Cameron2010}},
                                \tablefoottext{f}{\cite{Johnson2015}} - other studies \citep[e.g.,][]{Collier-Cameron2010} report larger uncertainties; hence, we have also performed our calculations with a $\varv_\mathrm{rot}\sin{i_*}$ that deviates by $\pm$\,10\,km\,s$^{-1}$, but could not find any significant differences in the results.     
                        }
                \end{threeparttable}
        \end{table}

Ultra-hot Jupiters are prime targets for atmospheric characterization by observing their thermal emission signal, given the elevated dayside temperatures of these objects. Various techniques have been used to study the emission signals from UHJ atmospheres, such as space-based photometric light curves \citep[e.g.,][]{Zhang2018, vonEssen2020, Deline2022}, low-spectral-resolution observations \citep[e.g.,][]{Arcangeli2018, ChangeatEdwards2021}, and ground-based high-resolution spectroscopy. In particular, emission spectroscopy has proven to be a powerful tool for identifying thermal inversions in the atmospheres of UHJs \citep[e.g.,][]{Haynes2015, Nugroho2017, Kreidberg2018, Yan2020, Kasper2021}. In these atmospheres, strong absorption of visible and ultraviolet radiation from the host star causes the temperature to increase with altitude. The presence of TiO and VO was initially proposed to cause this heating mechanism in upper planetary atmospheres \citep{Hubeny2003, Fortney2008}. However, theoretical work and more recent observations at high spectral resolution suggest that atomic metal species may also be fundamental for maintaining thermal inversion layers \citep[e.g.,][]{Arcangeli2018, Lothringer2018}. Neutral Fe is the species most commonly detected in thermal inversion layers, and is therefore thought to play an important role in the temperature structure of UHJ atmospheres \citep{Pino2020, Nugroho2020_Fe, Yan2020, Yan2021, Kasper2021, Cont2021, Herman2022}.

Another significant result of UHJ high-spectral-resolution observations is the frequent nondetection of Ti, V, and their oxides, although their signature should be present in the planetary spectra when assuming equilibrium chemistry \citep[e.g.,][]{Merritt2020, Hoeijmakers2020_WASP-121b, Tabernero2021, Yan2022}. This result suggests that depletion mechanisms may limit the presence of these species in the upper atmospheres of UHJs. Theoretical work has explored different mechanisms that may remove Ti- and V-bearing species from the atmospheres of hot giant exoplanets. For example, \cite{Spiegel2009} proposed the depletion of Ti and V in the upper part of planetary atmospheres via gravitational settling of TiO and VO. Cold trapping of Ti- and V-bearing molecules on the planetary nightsides has been suggested as another possible depletion scenario \citep{Parmentier2013}. However, these theoretical studies are mostly limited to the temperature range below that of UHJs. Further theoretical and observational work is therefore needed to better understand the circumstances under which Ti- and V-bearing species are depleted in UHJ atmospheres and how this impacts the planetary energy balance.

\begin{table*}
        \caption{Observation log and instrument characteristics.}             
        \label{obs_log}      
        \centering                          
        \renewcommand{\arraystretch}{1.3} 
        \begin{threeparttable}
                \begin{tabular}{l l l l l l l l l }        
                        \hline\hline                 
                        \noalign{\smallskip}
                        Instrument &  Spectral resolution ($R$) & Wavelength range ($\AA$) & Date & Phase coverage & Exposure time (s) & $N_\mathrm{spectra}$  \\     
                        \noalign{\smallskip}
                        \hline                       
                        \noalign{\smallskip}
                        CARMENES & 94\,600 (VIS) & 5200--9600    &  2017-11-15 &  0.29--0.65  & 300 & 95\tablefootmark{a}\\ 
                        & 80\,400 (NIR) & 9600--17\,100 &  2017-11-15 &  0.29--0.63  & 300 & 88\tablefootmark{a}\\ 
                        \noalign{\smallskip} \hline \noalign{\smallskip}
                        HARPS-N  & 115\,000 & 3830--6900 &  2020-10-15 &  0.43--0.70  & 200 & 125\\  
                        &  &  & 2020-11-07 &  0.24--0.57  & 200 & 155\\
                        \noalign{\smallskip} \hline \noalign{\smallskip}
                        ESPaDOnS & 68\,000 & 3700--10\,500 &  2013-09-15 &   0.30--0.44  & 90 & 110\\  
                        &  &  &   2013-09-26 &  0.37--0.44  & 90 & 55\\  
                        &  &  &   2014-09-04 &   0.56--0.69  & 90 & 110\\  
                        &  &  &   2014-09-15 &   0.55--0.68  & 90 & 110\\  
                        &  &  &   2014-11-05 &  0.31--0.38  & 90 & 55\\                          
                        \noalign{\smallskip}
                        \hline                                   
                \end{tabular}
                \tablefoot{
                        \tablefoottext{a}{Total number of CARMENES spectra was 105; due to their poor quality ten spectra were removed from the time series of the VIS channel and 17 spectra were removed from the time series of the NIR channel.}} 
        \end{threeparttable}      
\end{table*}
        
Several UHJs have been characterized using high-resolution Doppler spectroscopy. This technique uses the Doppler-shift of a planet's orbital motion relative to the telluric and stellar lines to identify its spectral signature \citep[e.g.,][]{Snellen2010, Brogi2012}. However, constraining the atmospheric parameters from the observed spectra remains a difficult task. For example, the elemental abundances and ratios or the atmospheric temperature structure cannot be accurately determined using high-resolution Doppler spectroscopy alone. To overcome this difficulty, atmospheric retrieval frameworks have been developed to fit high-resolution spectroscopy observations with parameterized model spectra. These retrievals result in statistical estimates of the physical parameters in the exoplanet atmosphere. Most of these retrievals rely on a likelihood function that compares the observed data to a forward model. A Bayesian estimator is then used to optimize the forward model parameters \citep[e.g.,][]{Brogi2017, BrogiLine2019, Yan2020, Gibson2020}. The use of alternative techniques for retrievals, such as supervised machine learning, has also been explored recently \citep{Fisher2020}.

In this study, we investigate the dayside emission spectrum of the UHJ WASP-33b. We report the first detection of atomic Ti and V via emission spectroscopy. Also, we present the physical parameters of the planetary atmosphere obtained from a Bayesian retrieval. WASP-33b orbits an A5-type star undergoing $\delta$ Scuti pulsations. Its bright host star ($V$\,$\sim$\,8\,mag), high equilibrium temperature ($T_\mathrm{eq}$\,$\sim$\,2700\,K), and short orbital period ($P_\mathrm{orb}$\,$\sim$\,1.22\,d) make WASP-33b a prime target for emission spectroscopy. Different studies have shown that the planet possesses a temperature inversion in its dayside hemisphere \citep[e.g.,][]{Haynes2015, Nugroho2017, Cont2021}. To date, the hydrogen Balmer lines, \ion{Si}{i}, \ion{Ca}{ii}, \ion{Fe}{i}, OH, CO, H$_2$O, and TiO have been identified via transmission or emission spectroscopy in the planetary atmosphere \citep{Nugroho2017, Nugroho2020_Fe, Nugroho2021, Yan2019, Yan2021, Cont2021, Cont2022, Herman2022, vanSluijs2022}. The WASP-33 system parameters used in this work are summarized in Table~\ref{tab-parameters}.

This work is organized as follows. We describe the observations and our data reduction routine in Sects.~\ref{Observations} and \ref{Data reduction}. The technique to search for individual species in the planetary atmosphere and the resulting detections are presented in Sect.~\ref{Detection of the planetary emission lines}. Our retrieval method and the inferred atmospheric parameters are described in Sect.~\ref{Atmospheric retrieval}. The conclusions of our work are given in Sect.~\ref{Conclusions}.

%

\section{Observations}
\label{Observations}    

The thermal phase curve of WASP-33b was observed at high spectral resolution over a total of eight visits. We observed the planetary emission spectrum on 15~November~2017 with CARMENES at the 3.5\,m Calar Alto telescope \citep{Quirrenbach2014, Quirrenbach2020}. Both the visible (VIS) and near-infrared (NIR) wavelength channels of CARMENES were used. We consider the CARMENES channels independently as two different instruments. We also observed the spectrum of WASP-33b during the two nights of 15~October~2020 and 7~November~2020 with HARPS-N at the Telescopio Nazionale Galileo \citep{Mayor2003, Cosentino2012}. Another five observations were obtained between September~2013 and November~2014 by \cite{Herman2020} with ESPaDOnS at the Canada-France-Hawai'i telescope \citep{Donati2003}. The combined set of nine observation datasets (two datasets from the CARMENES VIS and NIR observations, one dataset from each of the other seven observations) has already been used in previous studies to investigate the emission signature of TiO, \ion{Fe}{i}, and \ion{Si}{i} in the dayside atmosphere of WASP-33b \citep{Herman2020, Herman2022, Cont2021, Cont2022}. Details of the observations and the main characteristics of the used spectrographs are summarized in Table~\ref{obs_log}.

Prior to the data reduction, we discarded a number of spectra due to poor data quality. All the discarded spectra belong to the CARMENES observation night on 15~November~2017. For seven spectra, the star was not aligned with the telescope fiber, and for three spectra the flux drastically dropped due to a passing cloud. Another seven spectra were removed due to the insufficient flux level in the NIR channel caused by the low elevation of the target towards the end of the observation night (airmass\,>\,2). Consequently, we removed a total of 10 VIS channel spectra and 17 NIR channel spectra from the CARMENES observation. As no data quality issues were identified for the HARPS-N and ESPaDOnS observations, we included all spectra from the two instruments in the further analysis.

\section{Data reduction}
\label{Data reduction}

\subsection{Pre-processing the spectra}

The extraction of the one-dimensional spectra from the raw frames was performed with the data reduction pipelines of the respective instruments. We extracted the order-by-order spectra of CARMENES using the instrument pipeline {\tt caracal}~v2.20 \citep{Zechmeister2014, Caballero2016} and obtained the order-merged spectra of HARPS-N by running the {\tt Data Reduction Software} \citep{Cosentino2014}. The order-by-order ESPaDOnS spectra were extracted by the observatory using the {\tt Upena} pipeline, which relies on the data reduction package {\tt Libre-ESpRIT} \citep{Donati1997}. Except for the echelle orders 45--43 in the CARMENES NIR channel, we included the entire wavelength range of all instruments in our analysis. The three excluded CARMENES NIR orders coincide with the telluric absorption band of water at 1.4\,$\mu$m and therefore suffer from an insufficient flux level.

Each of the nine observation datasets was reduced separately. To normalize all the spectra to the same continuum level, we applied a polynomial fit to the individual spectra and divided them by the fit function. We used a second-order polynomial for the order-by-order spectra of CARMENES and ESPaDOnS and a seventh-order polynomial for the order-merged spectra of HARPS-N. Outliers were removed by applying a 5$\sigma$ clip to the time evolution of each pixel. Wavelength bins with the absorption level larger than 80\,\% of the spectral continuum were masked. We also masked the strong sky emission lines, which are present in the CARMENES NIR channel.

\subsection{Removal of telluric and stellar features}

We applied the detrending algorithm \texttt{SYSREM} to the spectral matrix of each observation to correct for the contribution of telluric and stellar lines. The algorithm was originally proposed for the removal of systematic effects from large sets of photometric light curves \citep{Tamuz2005}. \texttt{SYSREM} models the common systematic features of the light curves by iteratively fitting their trends as a function of time. Subsequently, the modeled systematics are subtracted from the data. When applied to the search for exoplanets, each wavelength bin of the spectral matrix is considered as an independent light curve. This procedure results in the so-called residual spectral matrix, which is the spectral matrix after removal of the systematics.

The input data for \texttt{SYSREM} are the spectral matrix and its uncertainties. For the CARMENES and ESPaDOns data, we used the propagated uncertainties from the instrument pipelines. As the HARPS-N pipeline does not compute uncertainties, we estimated them according to the procedure described by \cite{Yan2020}. This method consists of running \texttt{SYSREM} in a first step with five iterations and uniform uncertainty values. The resulting residual matrix is dominated by noise. We then calculated the average noise row by row and column by column. Finally, the uncertainty values of each data point are calculated as the mean of the respective row and column average noise.

We followed the approach from \cite{Gibson2020} and ran \texttt{SYSREM} in flux space instead of magnitude space \citep{Tamuz2005}. This method allows the relative strength of the planetary spectral lines to be preserved during the correction for the telluric and stellar contamination. First, we used the algorithm in the standard way to calculate a model of the systematic features. This model corresponds to the sum of the models from each \texttt{SYSREM} iteration. We then divided the model of the systematic features out of the original spectral matrix. Also, the uncertainties were divided by the model.

\begin{figure}
        \centering
        \includegraphics[width=0.5\textwidth]{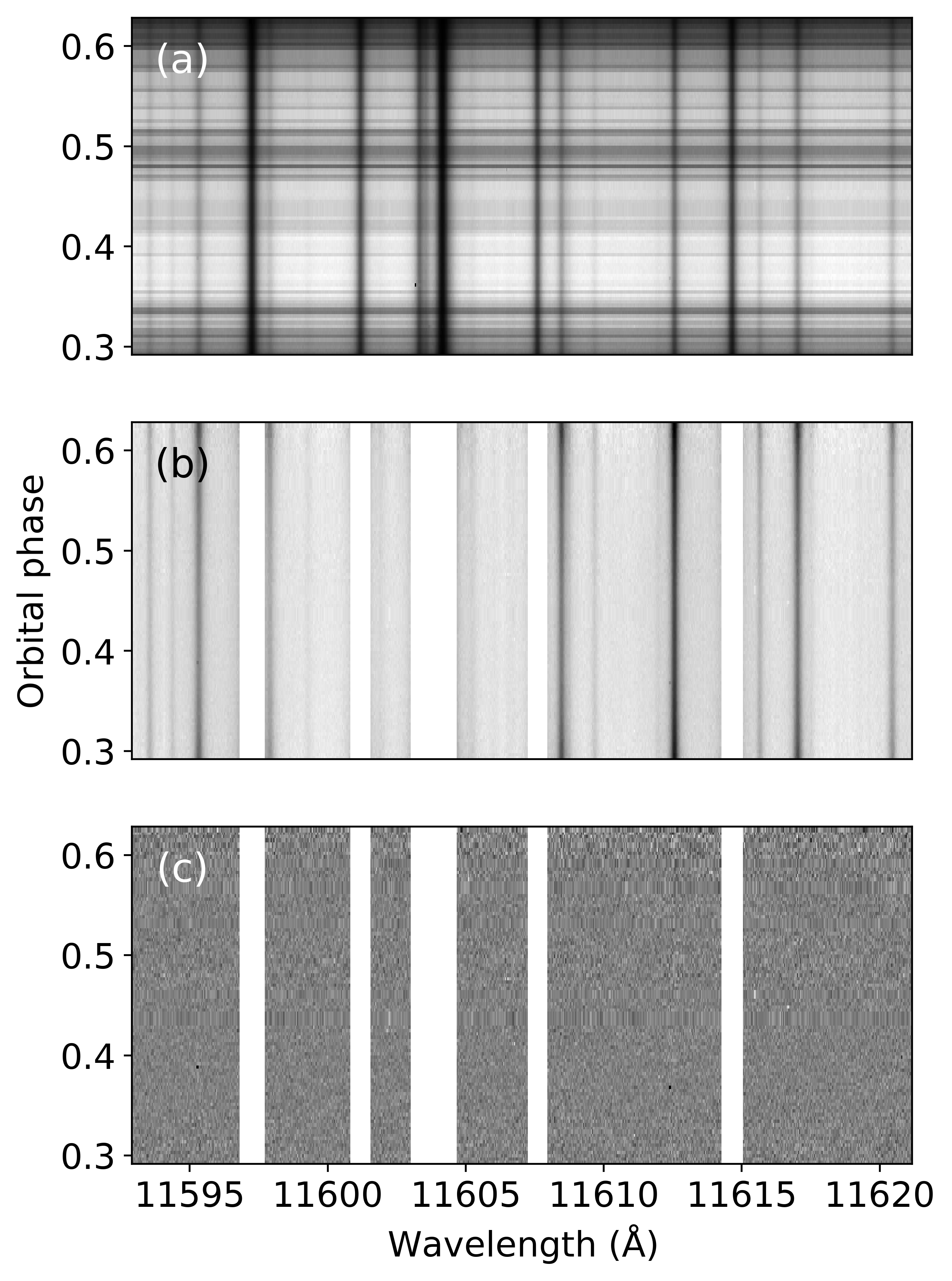}
        \caption{Data reduction steps for a representative CARMENES NIR wavelength range. {\it Panel~a} shows the unprocessed spectral matrix. {\it Panel~b} is the spectral matrix after normalization and outlier correction; we mask the strongest telluric lines in this step. {\it Panel~c} shows the spectra reduced with \texttt{SYSREM} after telluric and stellar line removal.}
        \label{SYSREM}
\end{figure}

We ran the algorithm over ten consecutive iterations. This results in a residual spectral matrix for each iteration. For the CARMENES and ESPaDOnS observations, we created the order-merged residual spectral matrix by combining the order-by-order \texttt{SYSREM} reduced spectral matrices. This step was not necessary for the HARPS-N data, because the spectra were already provided by the instrument pipeline in an order-merged format. An overview of the data reduction including \texttt{SYSREM} is provided in Fig.~\ref{SYSREM}.

\section{Detection of the planetary emission lines}
\label{Detection of the planetary emission lines}

We used the cross-correlation technique \citep[e.g.,][]{Snellen2010, Brogi2012, Sanchez-Lopez2019, Prinoth2022} to extract the weak emission signature of WASP-33b from the noise-dominated residual spectra. This method combines numerous spectral lines and translates them into a single peak via calculating the cross-correlation function (CCF) between the residual spectra and model spectra as described below. We searched for the emission lines of the metal species \ion{Ti}{i}, \ion{Ti}{ii}, \ion{V}{i}, and \ion{V}{ii}. Also, we aimed to confirm the presence of the hydroxyl radical (OH), which was recently detected by \cite{Nugroho2021}. Moreover, we reassessed the detections of \ion{Fe}{i} and \ion{Si}{i} in previous studies \citep{Cont2021, Cont2022} and investigated the presence of \ion{Fe}{ii} and \ion{Si}{ii} in the planetary atmosphere.

\subsection{Model spectra}
\label{Model spectra}

\begin{figure*}
        \centering
        \includegraphics[width=\textwidth]{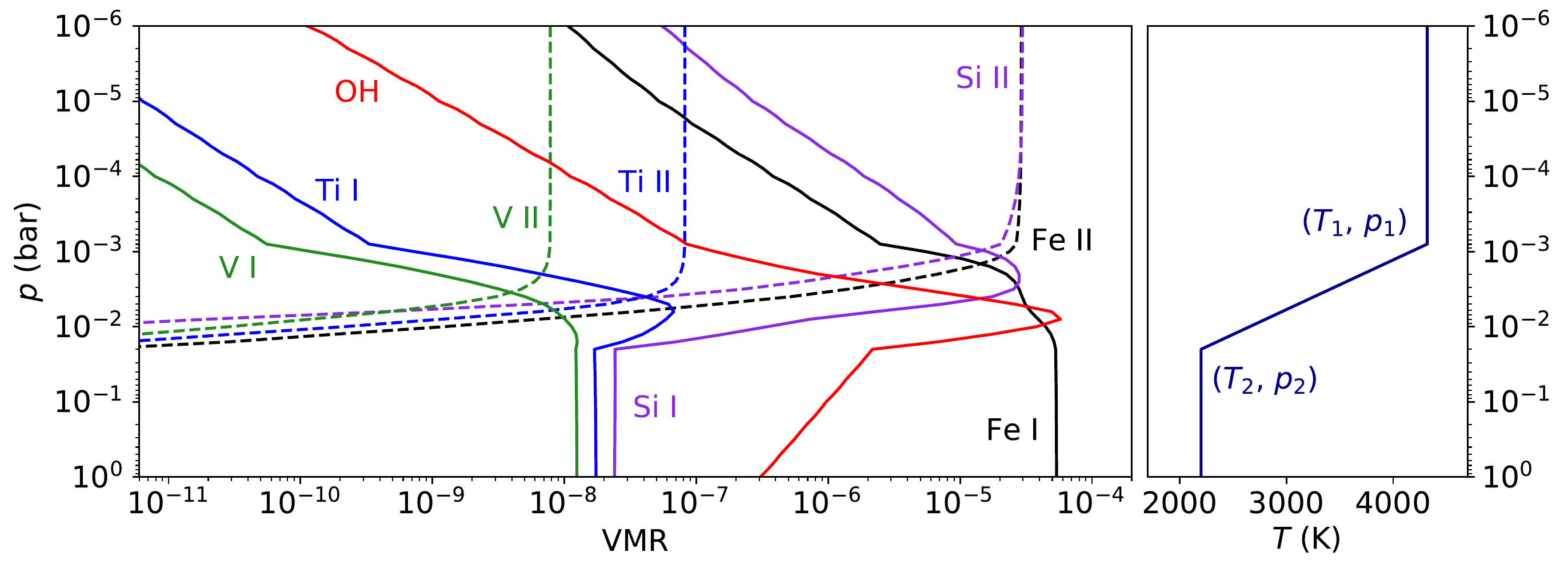}
        \caption{Volume mixing ratios (VMRs; {\it left panel}) and $T$-$p$ profile ({\it right panel}) used to generate the model spectra. We assumed equilibrium chemistry and solar elemental abundances for computing the VMRs. As the $T$-$p$ pattern is not known in detail, we assumed the two-point profile that was retrieved for WASP-189b by \cite{Yan2020}.}
        \label{VMR_and_TP}
\end{figure*}

We modeled a planetary atmosphere with 61 layers equispaced in log pressure from 1 to $10^{-6}$\,bar. As the atmospheric temperature profile of WASP-33b is not yet known in detail, we adopted the $T$-$p$ profile of WASP-189b, which was measured by \cite{Yan2020} via the \ion{Fe}{i} emission signature (Fig.~\ref{VMR_and_TP}). We believe that using the $T$-$p$ profile of WASP-189b is a reasonable approximation because the planet has similar properties to WASP-33b. This temperature profile has also been used successfully in previous work to characterize the atmosphere of WASP-33b \citep{Cont2021, Cont2022}. The $T$-$p$ profile is parametrized by a low pressure point ($T_1$, $p_1$) and a high pressure point \mbox{($T_2$, $p_2$)}. An isothermal atmosphere is assumed at pressures below $p_1$ or higher than $p_2$. Between these two isothermal layers, we assumed a temperature that changes linearly as a function of $\log{p}$. We used \texttt{easyCHEM} \citep{Molliere2017} to calculate the mean molecular weight and the volume mixing ratios (VMRs) of the investigated chemical species (Fig.~\ref{VMR_and_TP}). For this purpose, we assumed equilibrium chemistry and a solar elemental abundance. An opacity grid of each species was computed for modeling the planetary spectra. The metal opacities were calculated from the Kurucz line list \citep{Kurucz2018} and the OH opacities were obtained from the MoLLIST line database \citep{Brooke2016, Yousefi2018, Bernath2020}. Eventually, we ran the radiative transfer code \texttt{petitRADTRANS} \citep{Molliere2019} to compute the model spectrum of each species.

\begin{figure*}
        \centering
        \includegraphics[width=\textwidth]{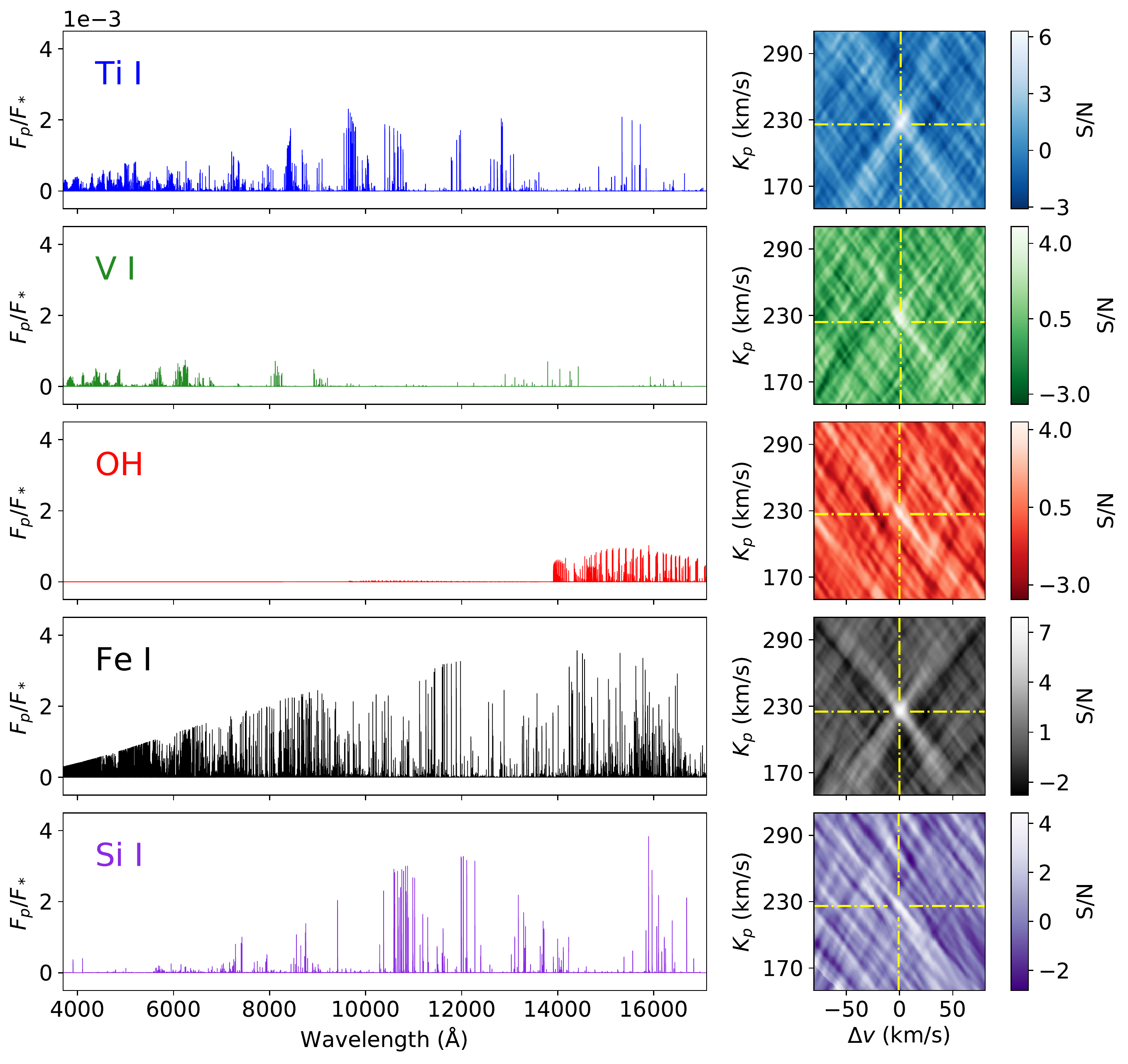}
        \caption{Model spectra and S/N maps of the neutral chemical species investigated. The \textit{left panels} show the normalized model spectra of the detected species. The presented interval 3700--17\,100\,$\AA$ corresponds to the combined wavelength coverage of the instruments used. The \textit{right panels} show the S/N maps of the neutral species that we searched for. For each species, the S/N map corresponds to the specific \texttt{SYSREM} iteration that maximizes the detection. The detection peaks are indicated by the yellow dash-dotted lines.}
        \label{SNmaps_and_template_spectra_neutral}
\end{figure*}

\begin{figure*}
        \centering
        \includegraphics[width=\textwidth]{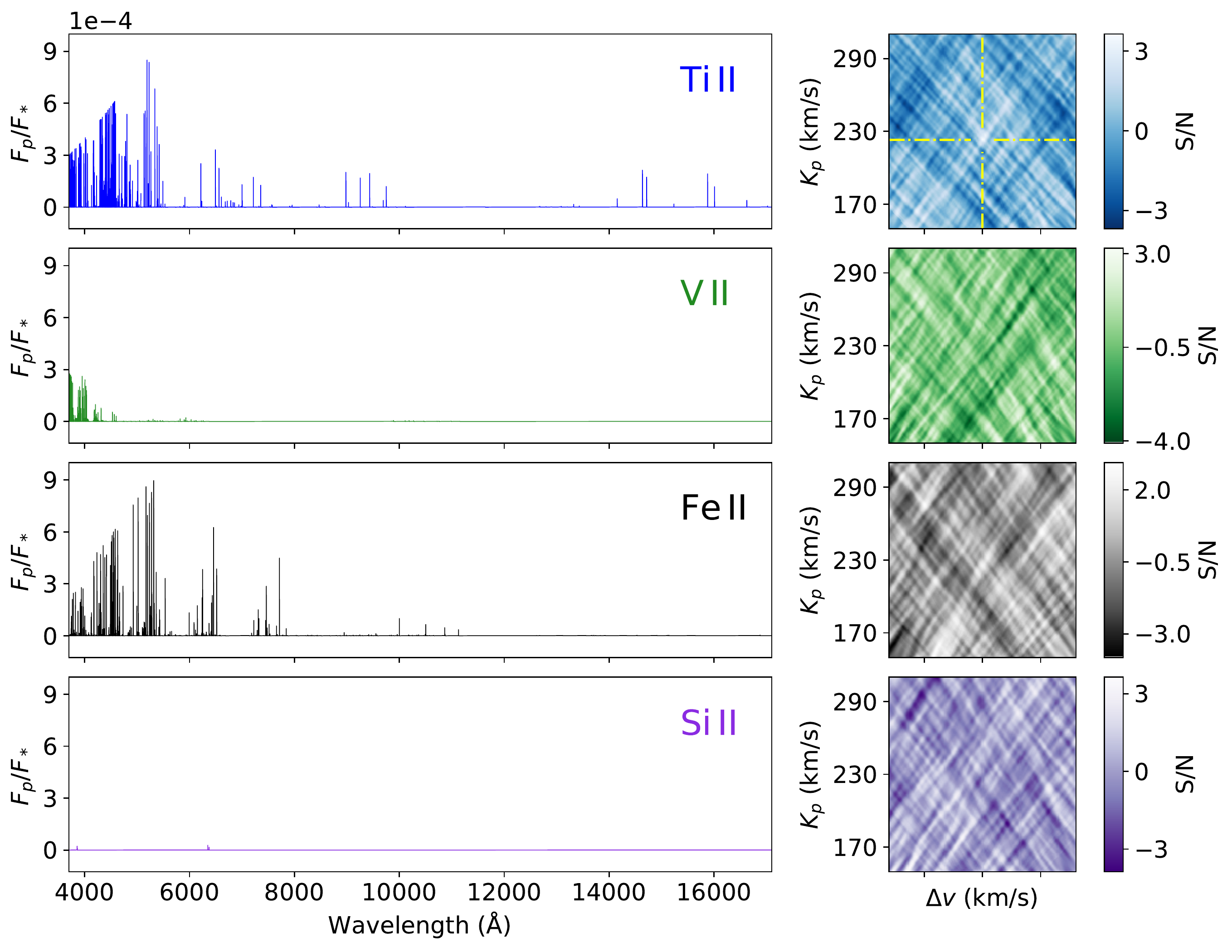}
        \caption{Same as Fig.~\ref{SNmaps_and_template_spectra_neutral}, but for the investigated ionic species. Only for \ion{Ti}{ii} a tentative detection peak that agrees with the planetary orbital motion is obtained. For \ion{V}{ii}, \ion{Fe}{ii}, and \ion{Si}{ii}, which are not detected at any \texttt{SYSREM} iteration, the S/N maps are the result of five consecutive iterations.}
        \label{SNmaps_and_template_spectra_ions}
\end{figure*}

As the reduced spectra were continuum normalized, a normalization of the model spectra was also required. First, we calculated the planet-to-star flux ratio of the model spectra by dividing with the blackbody spectrum of the host star. The result was then normalized to the planetary continuum. As a last step, we convolved each model spectrum with the instrument profiles, obtaining the final model spectra for cross-correlation. The normalized model spectra of all investigated species are illustrated in Figs.~\ref{SNmaps_and_template_spectra_neutral} and \ref{SNmaps_and_template_spectra_ions}.

\begin{table*}
        \caption{Summary of cross-correlation results.} 
        \label{tab-results} 
        \centering     
        \renewcommand{\arraystretch}{1.3} 
        \begin{threeparttable}
                \begin{tabular}{l l l l l l l}      
                        \hline\hline  
                        \noalign{\smallskip}
                        Species & S/N  & $K_\mathrm{p}$ (km\,s$^{-1}$)  & $\Delta \varv$ (km\,s$^{-1}$) & \texttt{SYSREM} iteration & Instruments included & Wavelength range ($\AA$) \\

                        \noalign{\smallskip}
                        \hline    
                        \noalign{\smallskip}
                        
                        \ion{Ti}{i}  & 6.3  & $226.0_{-3.5}^{+7.5}$ & $1_{-3}^{+5}$     & 3   & CV, CN, H, E  & 3700--17\,100  \\
                        \ion{V}{i}   & 4.8  & $224.0_{-3.0}^{+9.0}$ & $1_{-5}^{+2}$     & 3   & CV, H, E      & 3700--9600     \\
                        OH           & 4.4  & $227.0_{-23.0}^{+6.0}$ & $0_{-4}^{+18}$   & 3   & CN            & 9600--17\,100  \\
                        \ion{Fe}{i}  & 7.9  & $225.0_{-3.0}^{+3.5}$ & $0_{-2}^{+3}$     & 10  & CV, CN, H, E  & 3700--17\,100  \\
                        \ion{Si}{i}  & 4.4  & $226.0_{-17.5}^{+5.0}$ & $-1_{-10}^{+14}$ & 4   & CN            & 9600--17\,100  \\
                        \ion{Ti}{ii} & 3.6  & $223.0_{-2.5}^{+5.5}$ & $0_{-2}^{+4}$     & 1   & H             & 3830--6900     \\
                        \ion{V}{ii}  & \multicolumn{4}{l}{no detection}                       & H, E          & 3700--10\,500  \\
                        \ion{Fe}{ii} & \multicolumn{4}{l}{no detection}                       & H, E          & 3700--10\,500  \\
                        \ion{Si}{ii} & \multicolumn{4}{l}{no detection}                       & H, E          & 3700--10\,500  \\
                        \noalign{\smallskip}
                        \hline    
                        \noalign{\smallskip}                    
                        Combined     & 8.5  & $225.0_{-2.5}^{+3.0}$ & $0_{-2}^{+3}$     & 9  & CV, CN, H, E   & 3700--17\,100 \\
                        \noalign{\smallskip}
                        \hline                               
                \end{tabular}
                \tablefoot{For each species, only those instruments predicted to have prominent emission lines in their wavelength range were included. We use the following abbreviations: CARMENES VIS (CV), CARMENES NIR (CN), HARPS-N (H), and ESPaDOnS (E).}               
        \end{threeparttable}      
\end{table*}

\subsection{Cross-correlation method}
\label{Cross-correlation method}

The cross-correlation method was applied to each species independently. 
We Doppler-shifted the model spectrum from \mbox{--520\,km\,s$^{-1}$} to \mbox{+520\,km\,s$^{-1}$} with steps of 1\,km\,s$^{-1}$. Each order-merged residual spectrum was multiplied with the shifted model spectrum and weighted by the uncertainties. This resulted in the weighted CCF, defined as
\begin{equation}
\mathrm{CCF} = \sum r_i m_i(\varv).
\end{equation}
The data points of the residual spectra weighted by the inverse of the squared uncertainties are denoted by $r_i$, and $m_i$ is the model spectrum as a function of the Doppler-shift velocity $\varv$. By combining the CCFs from the different datasets in a two-dimensional array, we obtained the final CCF map. In this step, we included the CCFs of those instruments that cover the wavelength ranges with significant emission features of the species in consideration. In contrast, we did not include the data of the instruments for which the model spectra predicted insignificant spectral features. The instruments and wavelength ranges included in the analysis and the corresponding chemical species are listed in Table~\ref{tab-results}. 

\begin{figure}
        \centering
        \includegraphics[width=0.5\textwidth]{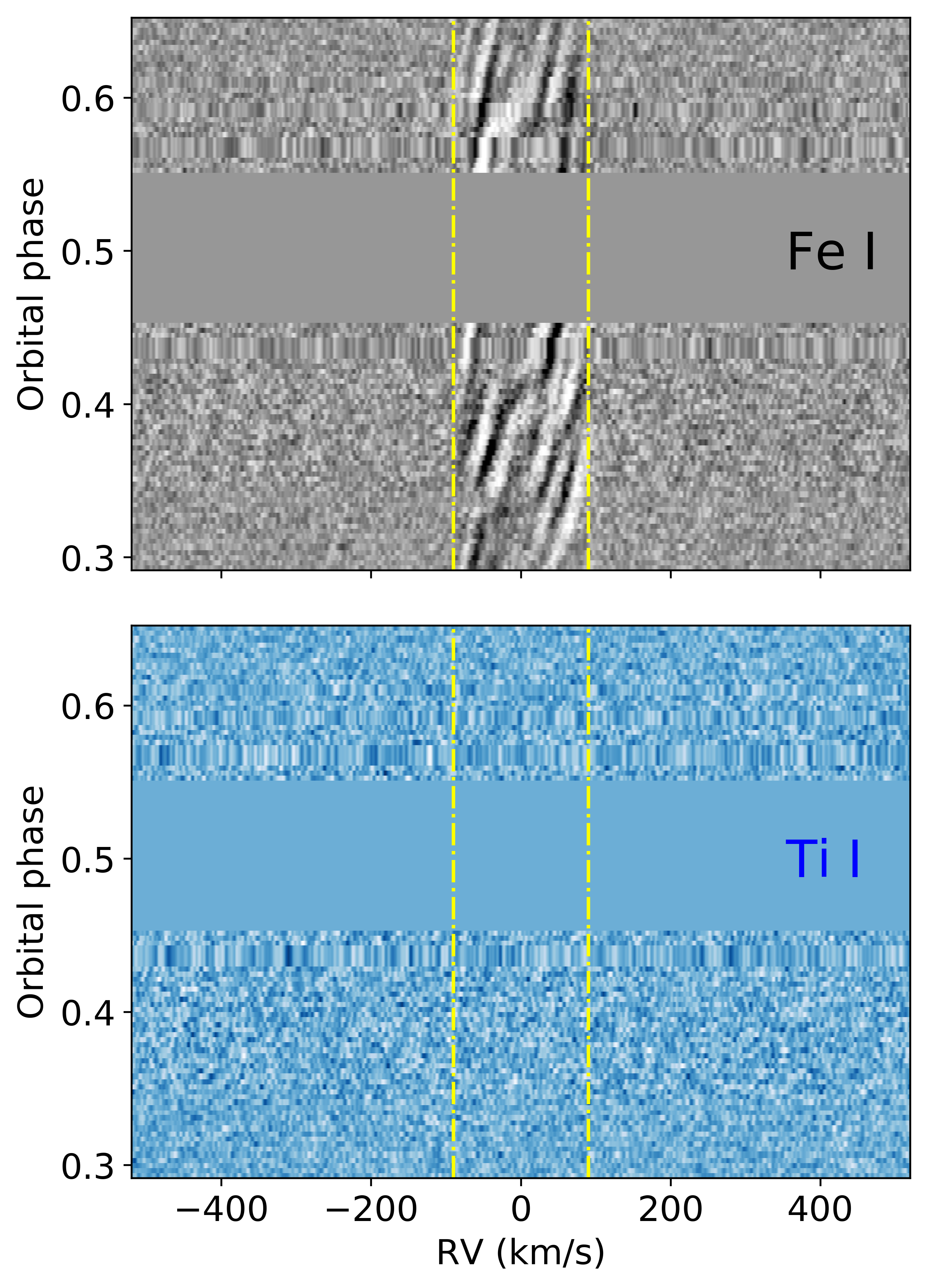}
        \caption{Example CCF maps from CARMENES VIS data. The RV region between $\pm \varv_\mathrm{rot}\sin{i_*}$ is indicated by the yellow dash-dotted lines. In the \textit{top panel} this region is dominated by artifacts from residual stellar \ion{Fe}{i} lines. Stellar artifacts are absent in the CCF map of \ion{Ti}{i} in the \textit{bottom panel}. The in-eclipse CCFs are masked in both panels.}
        \label{CCF_maps_comparison}
\end{figure}

WASP-33 is a $\delta$ Scuti variable star, showing time-dependent luminosity variations and a variable stellar line profile \citep{Herrero2011}. Given that \texttt{SYSREM} corrects only efficiently for stationary features, the variable stellar lines are not entirely removed by the algorithm from the observed spectra. The presence of residual stellar lines of the same species as the one under investigation can lead to artifacts in the CCF map. In this case, the affected radial velocity (RV) range in the CCF map depends on the stellar rotation velocity and lies between $\pm \varv_\mathrm{rot}\sin{i_*}$ in the stellar rest frame (i.e., $\pm$87\,km\,s$^{-1}$ for WASP-33). We find these artifacts to be present in the CCF maps of \ion{Ti}{ii}, \ion{V}{ii}, \ion{Fe}{i}, \ion{Fe}{ii}, \ion{Si}{i}, and \ion{Si}{ii}, and therefore masked the affected RV range in the CCF maps of these species. For \ion{Ti}{i}, \ion{V}{i}, and OH, we find no residual stellar lines present because there are no artifacts in their CCF maps, nor is a substantial change in the detections of these species with or without masking achieved. No masking is therefore applied to the CCF maps of these three species. In Fig.~\ref{CCF_maps_comparison} we show two example CCF maps, one with artifacts, the other without.

We assumed a circular orbit for shifting the CCF map of each species to the planetary rest frame. For this purpose, the CCFs were Doppler-shifted with a planetary RV of
\begin{equation}
\label{equ-orb-v}
\varv_\mathrm{p} = \varv_\mathrm{sys} + \varv_\mathrm{bary} + K_\mathrm{p} \sin{2\pi\phi} + \Delta \varv,
\end{equation}
with $K_\mathrm{p}$ the orbital velocity semi-amplitude, $\varv_\mathrm{sys}$ the velocity of the WASP-33 system, $\varv_\mathrm{bary}$ the barycentric correction, $\Delta \varv$ the velocity deviation from the planetary rest frame, and $\phi$ the orbital phase. We repeated the alignment over a range of different $K_\mathrm{p}$ values. For each alignment, the CCF map was collapsed into a one-dimensional CCF by calculating the mean value over all out-of-eclipse CCFs. The one-dimensional CCFs were then combined in a two-dimensional array. This array was normalized by its standard deviation under exclusion of the region around the strongest signal peak. In this way, we obtained a signal-to-noise map (S/N map) of the investigated chemical species. If the spectral signature is present in the data, the S/N map shows a significant peak at the location of the expected $K_\mathrm{p}$ and close to $\Delta \varv$\,=0\,km\,s$^{-1}$.

\subsection{Cross-correlation results and discussion}
\label{Cross-correlation results and discussion}

\begin{figure}[h]
        \centering
        \includegraphics[width=0.5\textwidth]{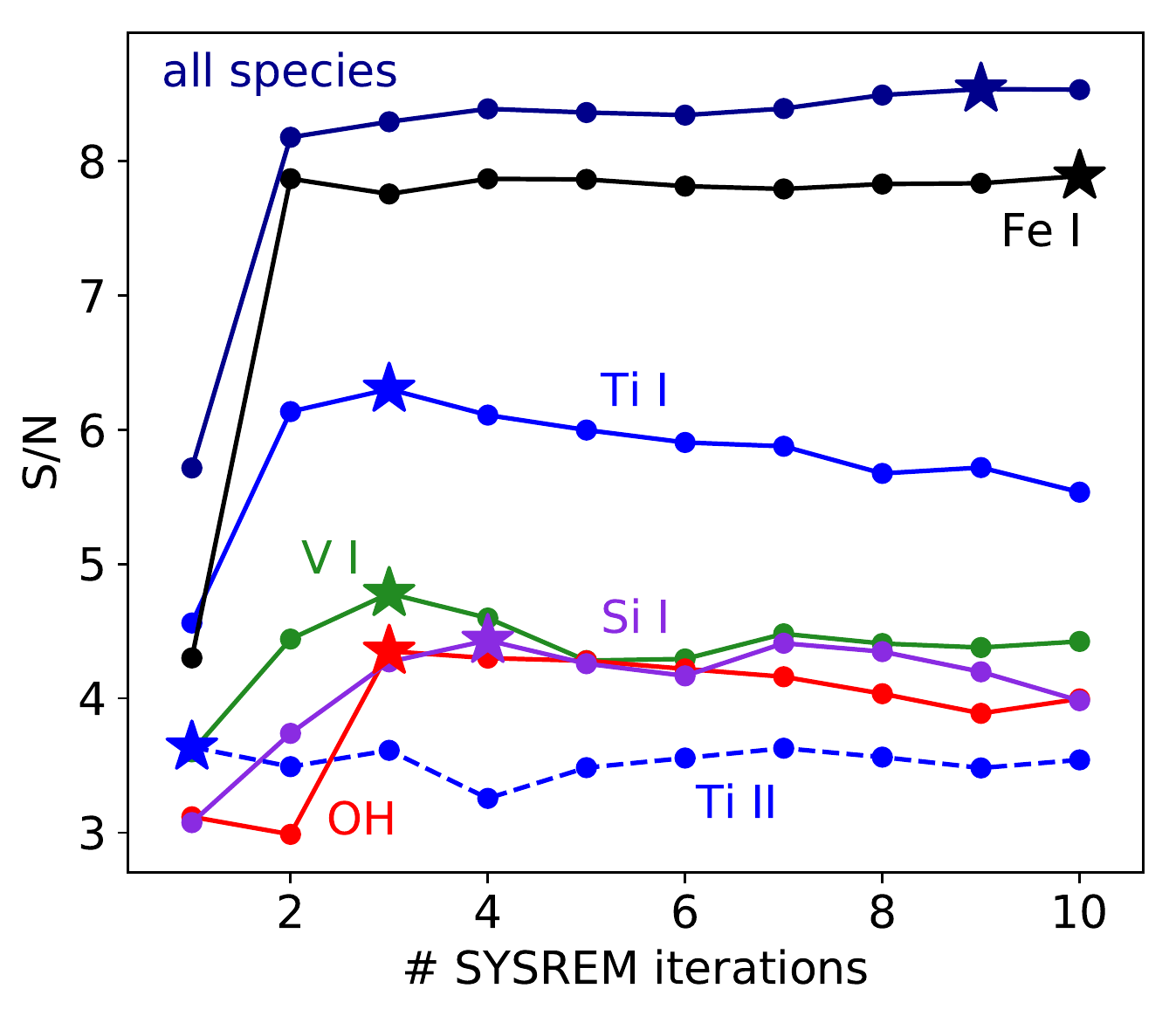}
        \caption{S/N detection strength as a function of \texttt{SYSREM} iteration. We show the pattern for \ion{Ti}{i}, \ion{V}{i}, OH, \ion{Fe}{i}, \ion{Si}{i}, the tentatively detected signature of \ion{Ti}{ii}, and the spectral feature of all species together. The iteration with the most significant S/N peak is indicated by the star symbol.}
        \label{SN_iterations}
\end{figure}

\begin{figure}
        \centering
        \includegraphics[width=0.5\textwidth]{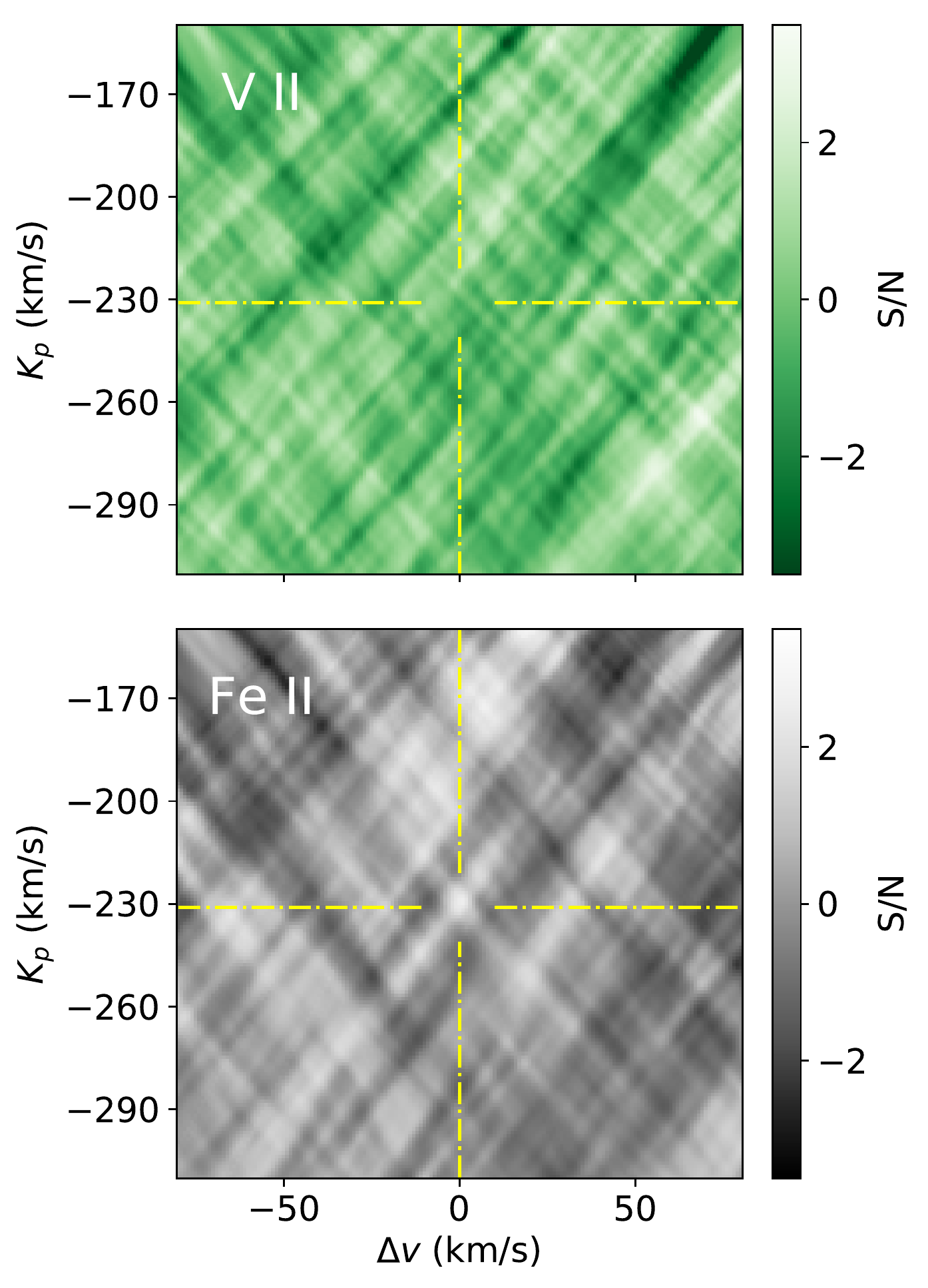}
        \caption{S/N maps of injection-recovery tests for \ion{V}{ii} and \ion{Fe}{ii}. The position of the injected signal is indicated by the yellow dashed-dotted lines. The injected \ion{V}{ii} signal cannot be detected, and that of \ion{Fe}{ii} causes a negligible peak in the S/N map. We conclude that the two species are not detectable even if they are present in the planetary atmosphere, given that the injected signals are poorly recovered.}
        \label{SN_injection-recovery_FeII_VII}
\end{figure}

\begin{figure*}
\centering
    \includegraphics[width=\textwidth]{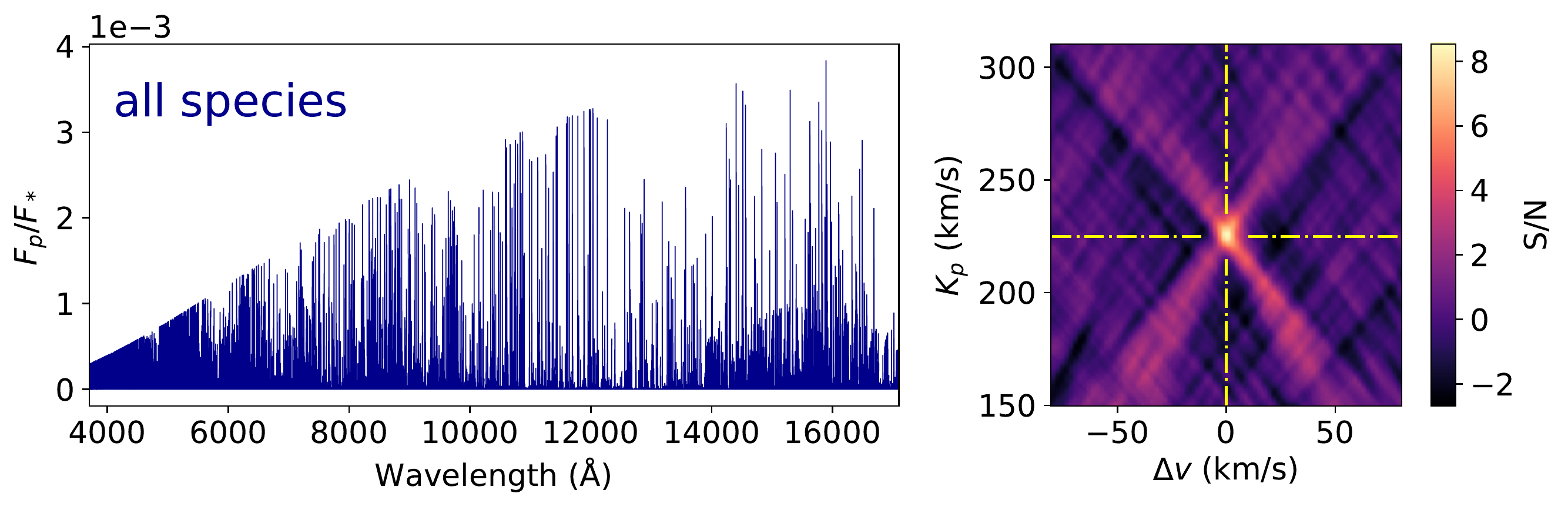}
\caption{Model spectrum and detected signature from combining the emission lines of all the detected chemical species. The \textit{left panel} is the normalized model spectrum that includes the spectral lines of \ion{Ti}{i}, \ion{V}{i}, OH, \ion{Fe}{i}, \ion{Si}{i}, and \ion{Ti}{ii}. In the \textit{right panel}, we show the resulting S/N map. The detection peak coordinates are indicated by the yellow dash-dotted lines.}
\label{SN_model_all-species}
\end{figure*}

We find significant detections ($>4\sigma$) of \ion{Ti}{i}, \ion{V}{i}, OH, \ion{Fe}{i}, and \ion{Si}{i} that correspond to the Doppler-velocity of WASP-33b's orbital motion. A clear peak is seen for all \texttt{SYSREM} iterations greater than one in the S/N maps at the location of the expected orbital parameters (Fig.~\ref{SN_iterations}). Figure~\ref{SNmaps_and_template_spectra_neutral} shows the S/N maps and the corresponding model spectra used for cross-correlation. We summarize the S/N of the detections and the measured orbital parameters in Table~\ref{tab-results}.

This is the first detection of \ion{Ti}{i} and \ion{V}{i} in the atmosphere of WASP-33b. Previously, these two species were detected exclusively in transmission, making this their first detection in an exoplanet atmosphere using emission spectroscopy \citep{BenYami2020, Stangret2022, Ishizuka2021, Gibson2022, Kesseli2022, BelloArufe2022, Prinoth2022}. 
Moreover, our results confirm the recent report of OH in the dayside atmosphere of WASP-33b \citep{Nugroho2021}. 
The detections of \ion{Fe}{i} and \ion{Si}{i} are a re-evaluation of earlier work \citep{Cont2021, Cont2022} and are shown here for completeness. We note that the orbital parameters found for the individual species differ slightly, but within the one-sigma range. Detecting the emission signature from multiple species also unambiguously proves the existence of a temperature inversion layer in the planetary atmosphere, which is in agreement with the predictions from theoretical work \citep{Hubeny2003, Fortney2008, LothringerBarman2019}.

Figure~\ref{SNmaps_and_template_spectra_ions} illustrates the cross-correlation results for the ionic species \ion{Ti}{ii}, \ion{V}{ii}, \ion{Fe}{ii}, and \ion{Si}{ii}. Emission lines of \ion{Ti}{ii}, \ion{Fe}{ii,} and, to a lesser extent, \ion{V}{ii} are expected to be present near the blue end of the observed wavelength range. Only in the specific case of \ion{Ti}{ii} and when considering the data from HARPS-N alone do we find tentative evidence for emission lines consistent with the planetary rest frame. Given the strong detection of \ion{Ti}{i} and the higher spectral resolution of HARPS-N in comparison to the other instruments used, we consider the tentative \ion{Ti}{ii} peak as a real signal. On the other hand, no significant detections of \ion{V}{ii} and \ion{Fe}{ii} are achieved. We ran an injection-recovery test to investigate the nondetections of the two species. To this end, we Doppler-shifted the convolved model spectra used to attempt the detection of \ion{V}{ii} and \ion{Fe}{ii} with the reversed $K_\mathrm{p}$ of WASP-33b \citep{Merritt2020}. Subsequently, we injected the shifted model spectra into the raw data and performed the pre-processing and cross-correlation analysis as described in Sects.~\ref{Data reduction} and \ref{Cross-correlation method}. Doppler-shifting the injected signal with the reversed $K_\mathrm{p}$ value of --231\,km\,s$^{-1}$ prevents interference with potentially undetected \ion{V}{ii} and \ion{Fe}{ii} signals from the planetary atmosphere. No recovery could be achieved for \ion{V}{ii}, and the injected \ion{Fe}{ii} model spectrum resulted in a negligible peak in the S/N map. We show the results of the injection-recovery test in Fig.~\ref{SN_injection-recovery_FeII_VII}. Given the poor recovery of the injected \ion{V}{ii} and \ion{Fe}{ii} signals, we conclude that our nondetections of these species are likely due to the relatively small number of prominent emission lines rather than resulting from their absence in the planetary atmosphere. Our nondetection of the \ion{Si}{ii} signal was expected because the emission lines are very weak over the considered wavelength range. In summary, obtaining a cross-correlation signal of the metal ions studied is more challenging than detecting the corresponding neutral species because of the smaller number and strength of their emission lines, which are restricted to a relatively narrow wavelength range.

Our results show that atomic Ti and V are not rained out, cold trapped, or otherwise depleted in the hot atmosphere of WASP-33b. Moreover, the detections of \ion{Ti}{i}, \ion{V}{i}, and the tentative signal of \ion{Ti}{ii} are consistent with the identification of these species in the atmospheres of a number of other strongly irradiated exoplanets \citep[e.g.,][]{Hoeijmakers2018, Stangret2022, Prinoth2022, BelloArufe2022, Kesseli2022}. We therefore conclude that the presence of significant Ti and V concentrations is favored at elevated temperatures. Because emission spectroscopy preferentially probes spectral lines emerging from the hot planetary dayside, we propose this method to be particularly suitable for the search for refractory species such as Ti and V in exoplanet atmospheres. However, some observations suggest that the presence or absence of refractory species may not be determined by the atmospheric temperature alone. For example, \ion{Ti}{i} was detected in the planetary atmosphere of HD~149026b \citep{Ishizuka2021}, the temperature of which is significantly below that of UHJs. On the other hand, the signature of the same species was not identified in the very hot atmosphere of KELT-20b/MASCARA-2b \citep{Yan2022}. These observations suggests that physical parameters other than the temperature may also be important for the presence of Ti and V in the atmospheres of gas giant exoplanets.

We further applied the cross-correlation method to the datasets with a model spectrum that includes all the detected species (i.e., \ion{Ti}{i}, \ion{V}{i}, OH, \ion{Fe}{i}, \ion{Si}{i}, and \ion{Ti}{ii}). The S/N map and the model spectrum are shown in Fig.~\ref{SN_model_all-species}. We find that using the model spectrum of all the species together results in an overall detection strength of S/N\,=\,8.5 after eight \texttt{SYSREM} iterations (Fig.~\ref{SN_iterations}). A comparison with the S/N values from the individual species in Table~\ref{tab-results} shows that the spectral signature of \ion{Fe}{i} is driving the detection. This is a reasonable finding, because the \ion{Fe}{i} lines are expected to dominate the planetary emission spectrum over the entire wavelength range of our observations. The retrieved orbital parameters result in $K_\mathrm{p}$ and $\Delta \varv$ values of $225.0_{-2.5}^{+3.0}$\,km\,s$^{-1}$ and $0_{-2}^{+3}$\,km\,s$^{-1}$, respectively. Our $K_\mathrm{p}$ is slightly lower than the expected value of $231\pm3$\,km\,s$^{-1}$ \citep{Kovacs2013, Lehmann2015}, but is consistent with the results of previous studies \citep{Nugroho2020_Fe, Cont2021, Cont2022}. The retrieved $\Delta \varv$ value is consistent with the orbital motion of the planet. This indicates that the planet may not have a strong dayside to nightside wind at the altitudes probed by these emission lines, which would cause a deviation of the detection peak of the order of a few km\,s$^{-1}$ from the planetary rest frame.

\begin{figure*}[h]
\centering
\begin{subfigure}[c]{0.5\textwidth}
        \centering
        \includegraphics[width=\textwidth]{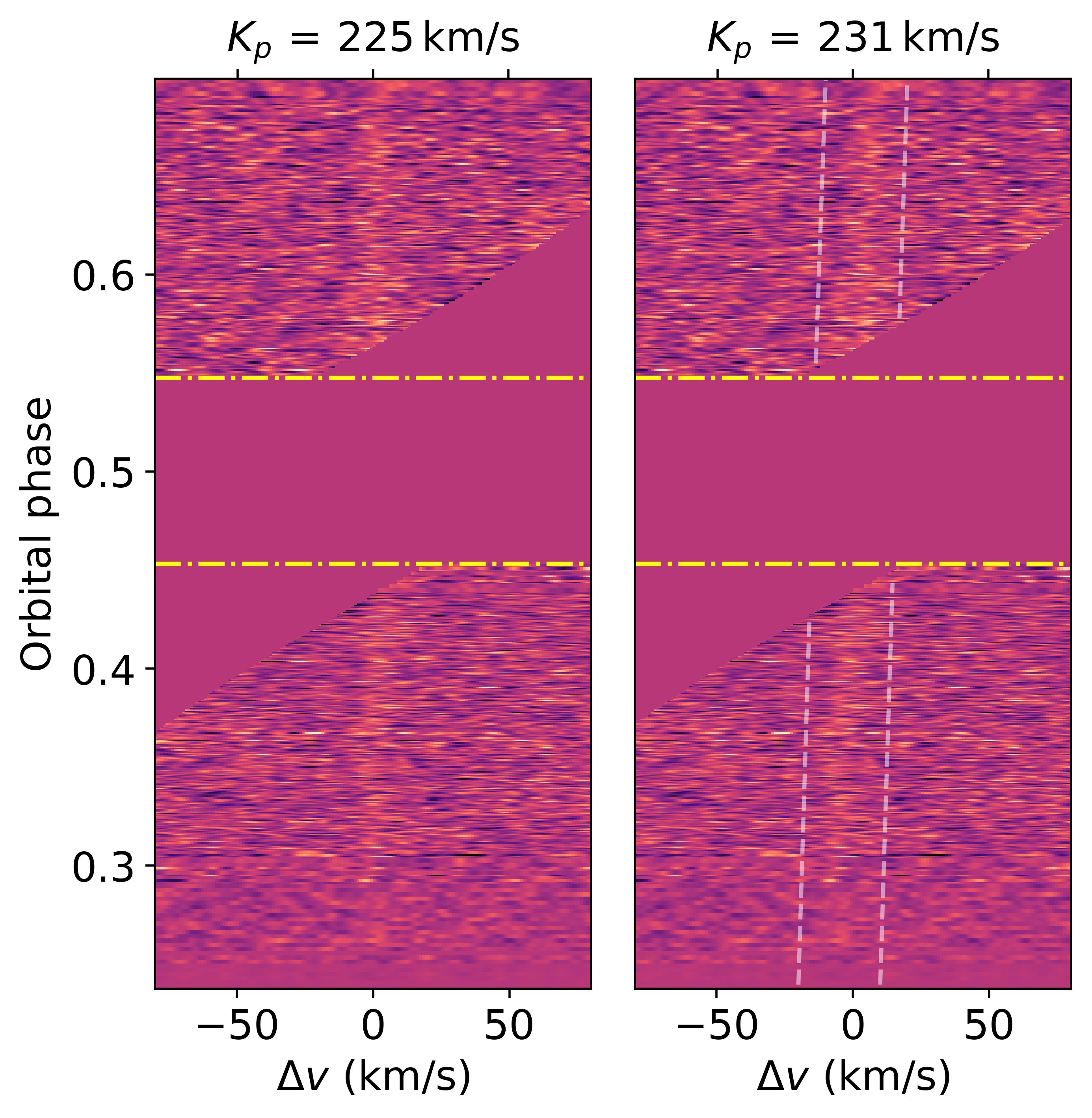}
\end{subfigure}
\quad
\begin{subfigure}[c]{0.4\textwidth}  
        \centering 
        \includegraphics[width=\textwidth]{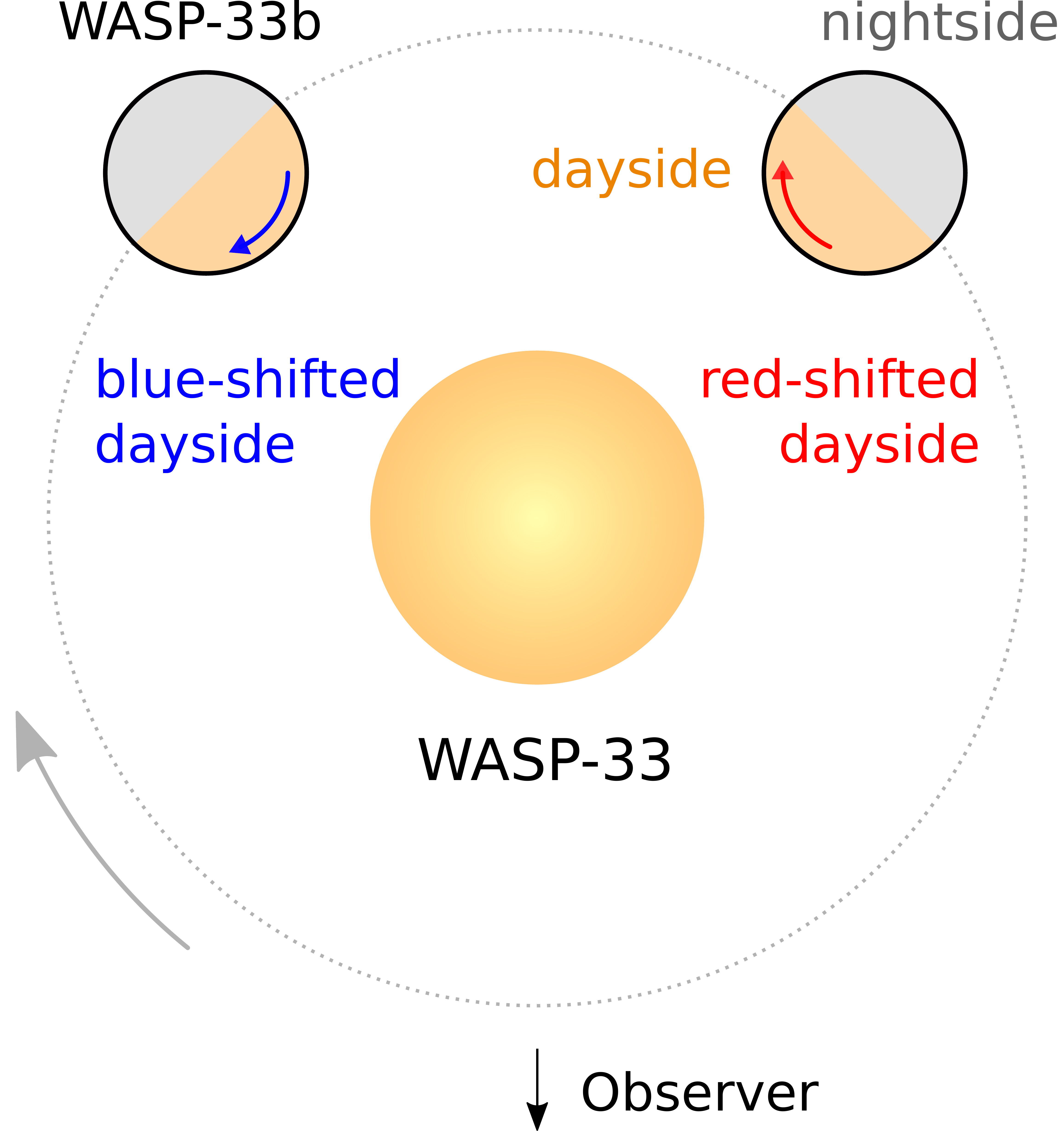}
\end{subfigure}
\vskip\baselineskip
\caption{CCF maps including all the detected chemical species and Doppler-shift induced by planetary rotation. The two CCF maps are aligned using the $K_\mathrm{p}$ values of $225$\,km\,s$^{-1}$ (\textit{left panel}; detected $K_\mathrm{p}$) and $231$\,km\,s$^{-1}$ (\textit{middle panel}; expected $K_\mathrm{p}$), respectively. When shifting the CCFs with the detected $K_\mathrm{p}$, no offset from the planetary rest frame is found. The alignment with the expected $K_\mathrm{p}$ yields a blueshift before and a redshift after secondary eclipse, which is indicated by the white dashed lines. We indicate the secondary eclipse with the yellow dash-dotted lines. The masked regions in the CCF map outside eclipse correspond to the location of residual stellar lines. The \textit{right panel} illustrates the WASP-33 system. The curved blue and red arrows indicate the planetary rotation direction, and the gray arrow indicates the orbital motion. Due to the planetary rotation, the signature from the planetary dayside experiences an additional blueshift before eclipse and redshift after eclipse.} 
\label{CCFs_225_231_all-species}
\end{figure*}

Figure~\ref{CCFs_225_231_all-species} shows two versions of the CCF map, one aligned with the detected $K_\mathrm{p}$ of $225$\,km\,s$^{-1}$, the other with the expected value of $231$\,km\,s$^{-1}$. The emission signal aligned with the detected $K_\mathrm{p}$ value appears as a vertical trail with zero offset from the planetary rest frame. However, aligning the CCFs with the expected $K_\mathrm{p}$ results in a tilted trail that is blue- and redshifted by a few km\,s$^{-1}$ before and after secondary eclipse, respectively. Assuming that the expected $K_\mathrm{p}$ reflects the true orbital velocity of the planet, we propose that these Doppler shifts are caused by the fast rotation velocity of WASP-33b of $\sim7$\,km\,s$^{-1}$ and the possible presence of super-rotation. In this scenario, the signature of the dayside atmosphere undergoes a spectral blueshift before eclipse and a redshift after eclipse (Fig.~\ref{CCFs_225_231_all-species}).

\section{Retrieval of the atmospheric properties}
\label{Atmospheric retrieval}

We used the retrieval method developed by \cite{Yan2020} to constrain the properties of WASP-33b's atmosphere. This method has already been successfully used to determine the thermal structure in the atmosphere of two other UHJs, WASP-189b and KELT-20b/MASCARA-2b \citep{Yan2020, Yan2022, Borsa2022}. In our implementation, we further developed this method for use with data from multiple instruments with different wavelength coverage.

\subsection{Retrieval method}
\label{Retrieval method}

\begin{figure*}
        \centering
        \includegraphics[width=\textwidth]{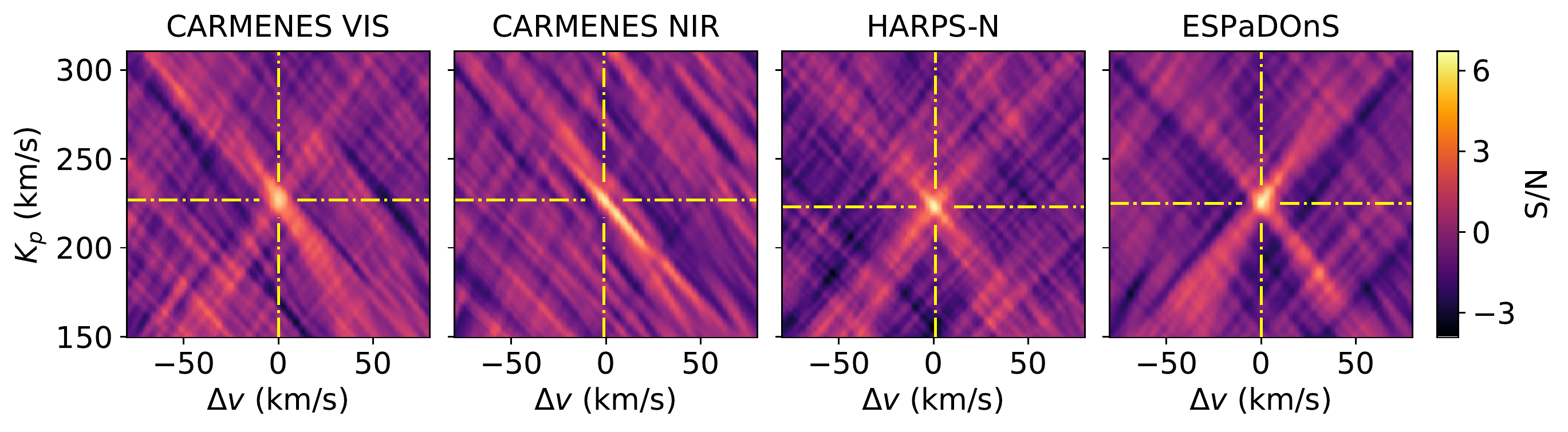}
        \caption{Instrument-specific S/N maps including all the detected chemical species (i.e., \ion{Ti}{i}, \ion{V}{i}, OH, \ion{Fe}{i}, \ion{Si}{i}, and \ion{Ti}{ii}). The orbital parameters and the optimal number of \texttt{SYSREM} iterations from this analysis are used for aligning the spectra. The detection peaks are indicated by the yellow dash-dotted lines.}
        \label{SNmaps_all_species}
\end{figure*}

First, we calculated an individual master residual spectrum for each instrument. We used Eq.~\ref{equ-orb-v} to shift the residual spectra to the planetary rest frame. For aligning the spectra, we used the values of $K_\mathrm{p}$ and $\Delta \varv$ measured for each instrument individually. The measurements were performed using the spectral model that consists of all the detected species. The corresponding S/N maps are reported in Fig.~\ref{SNmaps_all_species}; the orbital parameters and \texttt{SYSREM} iterations used are summarized in Table~\ref{tab-all-species}. To obtain the master residual spectrum, we computed the average of the shifted residual spectra, weighted by the squared S/N of each exposure frame. As each spectrum corresponds to a different orbital phase value, the final result of our retrieval reflects the average atmospheric conditions over the observed phase interval. The master residual spectrum is still dominated by noise and contains the continuum normalized planet-to-star flux ratio.

\begin{table}
        \caption{Summary of instrument-specific results from cross-correlation.} 
        \label{tab-all-species} 
        \centering     
        \renewcommand{\arraystretch}{1.3} 
        \begin{threeparttable}
                \begin{tabular}{l l l l r}      
                        \hline\hline  
                        \noalign{\smallskip}
                        Instrument & S/N  & $K_\mathrm{p}$  & $\Delta \varv$ & \texttt{SYSREM} \\
                        & & (km\,s$^{-1}$)& (km\,s$^{-1}$) & iteration \\
                        
                        \noalign{\smallskip}
                        \hline    
                        \noalign{\smallskip}
                        
                        CARMENES VIS & 5.9  & $227.0_{-5.0}^{+5.0}$ & $0_{-3}^{+4}$     & 2 \\
                        CARMENES NIR & 6.4  & $227.0_{-18.0}^{+3.5}$ & $-1_{-3}^{+15}$     & 7    \\
                        HARPS-N      & 6.7  & $223.0_{-3.0}^{+3.5}$ & $1_{-3}^{+2}$   & 3           \\
                        ESPaDOnS     & 6.7  & $225.0_{-3.0}^{+5.5}$ & $0_{-2}^{+4}$     & 4 \\
                        
                        \noalign{\smallskip}
                        \hline                               
                \end{tabular}
                \tablefoot{We used the spectral lines of all the detected chemical species (i.e., \ion{Ti}{i}, \ion{V}{i}, OH, \ion{Fe}{i}, \ion{Si}{i}, and \ion{Ti}{ii}).}
        \end{threeparttable}      
\end{table}

In a second step, we defined the model spectrum for fitting with the master residual spectra from the four instruments. To this end, we used the radiative transfer code \texttt{petitRADTRANS} \citep{Molliere2019}. Our opacity grid covers a temperature range up to 25\,000\,K for all metal species. In the specific case of OH, the available partition function limited the opacity calculations to 5000\,K. All OH opacities above this temperature were approximated with the 5000\,K value. This limitation is caused by the fact that most of the OH molecules are thermally dissociated beyond this temperature. The planetary atmosphere was modeled with 25 layers equally spaced over a logarithmic scale between 1 and $10^{-8}$\,bar. A two-point $T$-$p$ parametrization, as described in Sect.~\ref{Model spectra}, was used. 
We used the chemical equilibrium code \texttt{easyCHEM} \citep{Molliere2017} to compute the VMRs as a function of the atmospheric elemental abundance, assuming that for all metals it varies with the overall metallicity. 
Moreover, we included the effect of spectral line broadening to account for the presence of atmospheric turbulence and bulk motion. To this end, we assumed a Gaussian as the line broadening function and set the standard deviation as $\varv_\mathrm{broad}$. We then convolved the model spectrum with the Gaussian function in velocity space to account for the line broadening. Following our description in Sect.~\ref{Model spectra}, the model spectrum was further converted to the planet-to-star flux ratio and convolved with the instrument profile. The model spectrum was then interpolated to the wavelength solution of the master residual spectrum of each instrument.

Finally, we fitted the spectral model to the master residual spectra. For each instrument, we used a standard Gaussian log likelihood function
\begin{equation}
\ln{L} = -\frac{1}{2}\sum_{i} \left[ \frac{(R_i - m_i)^2}{(\beta \sigma_i)^2} + \ln{2 \pi (\beta \sigma_i)^2} \right], 
\end{equation}
with $R_i$ the data points of the residual spectrum, $\sigma_i$ their uncertainties, $\beta$ a scaling term to correct for possible overestimation or underestimation of the uncertainties, and $m_i$ the spectral model. We co-added the functions of the different instruments to get the combined log likelihood function of all the data. This results in an independent noise scaling factor for each instrument. For evaluating the combined likelihood function and estimating the fit parameters, we ran the Markov Chain Monte Carlo (MCMC) method from the \texttt{emcee} software package \citep{ForemanMackey2013}. Our retrieval includes the following free parameters: the temperature profile $T_1$, $p_1$, $T_2$, $p_2$; a noise scaling parameter for each instrument $\beta_\mathrm{CV}$, $\beta_\mathrm{CN}$, $\beta_\mathrm{H}$, $\beta_\mathrm{E}$\footnote{Each instrument has an independent noise scaling factor. We use the following abbreviations: CARMENES VIS (CV), CARMENES NIR (CN), HARPS-N (H), and ESPaDOnS (E).}; the overall metallicity [M/H]; the average broadening velocity $\varv_\mathrm{broad}$. For each free parameter, we used 24 walkers with 4000 steps in the sampling.

\subsection{Retrieval results and discussion}

\subsubsection{Retrieval including all the detected species}
\label{Retrieval including all the detected chemical species}

\begin{figure*}
        \centering
        \includegraphics[width=\textwidth]{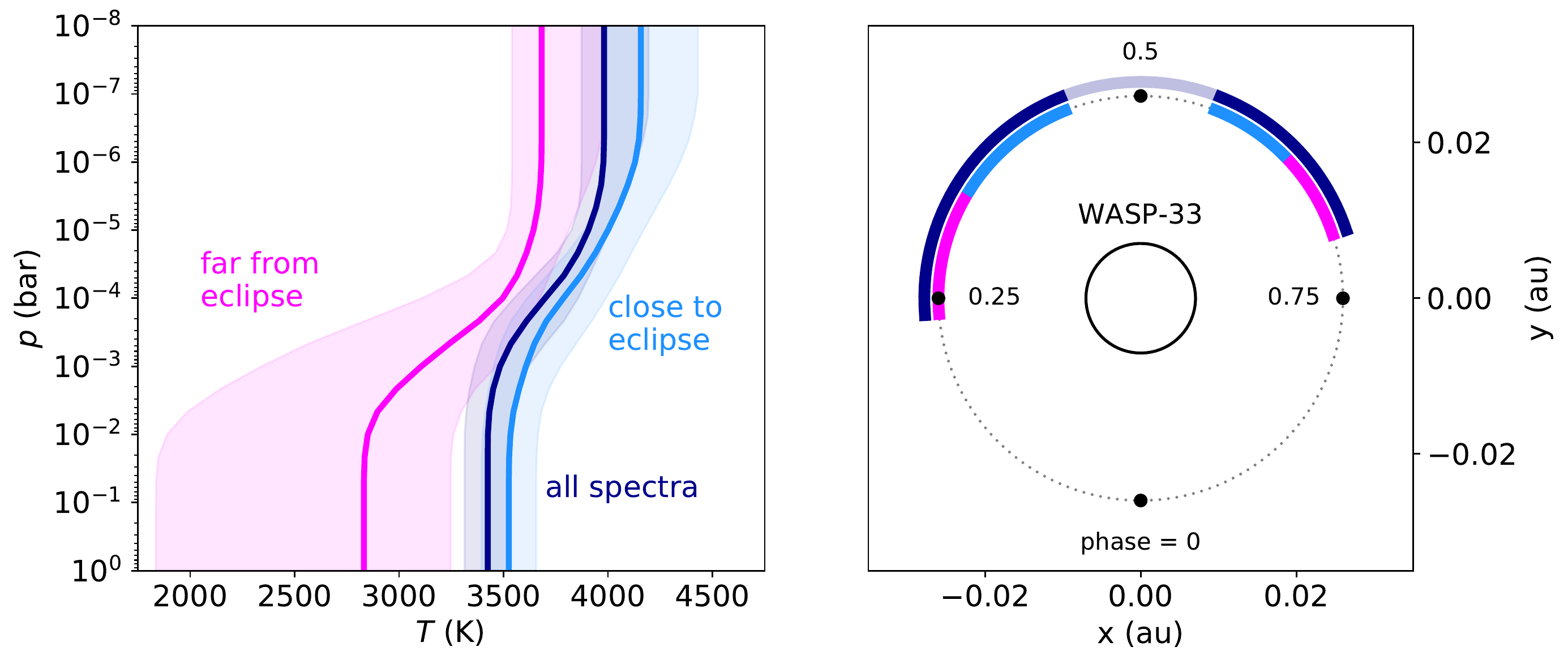}
        \caption{Temperature curves from the retrievals including all the detected chemical species and orbital phase coverage. The \textit{left panel} illustrates the median of the sampled temperature profiles with the 1$\sigma$ interval. The dark blue line indicates the temperature profile computed using all the spectra outside the RV interval $\pm \varv_\mathrm{rot}\sin{i_*}$. We also measured the temperature profiles by running the retrieval on two subsets of the spectral time series. The $T$-$p$ profile inferred from the spectra far from secondary eclipse is indicated in pink, while that inferred from the spectra close to eclipse is in light blue. The \textit{right panel} shows the orbital phase intervals used to compute the different $T$-$p$ profiles (orbital phases in the RV range $\pm \varv_\mathrm{rot}\sin{i_*}$ in transparent dark blue).}
        \label{retrieved-TP}
\end{figure*}

We included the emission lines of all the detected chemical species from Sect.~\ref{Detection of the planetary emission lines} into the retrieval (i.e., \ion{Ti}{i}, \ion{V}{i}, OH, \ion{Fe}{i}, \ion{Si}{i}, and \ion{Ti}{ii}). Only the spectra with a planetary Doppler velocity outside the RV interval $\pm \varv_\mathrm{rot}\sin{i_*}$ were used. In this way, we can be sure that the residual stellar lines do not overlap with the planetary spectral signature. The excluded spectra correspond to $\sim$10\% of the cumulative exposure time. Table~\ref{tab-retrieval-results} summarizes the best-fit parameters resulting from the retrieval. The posterior distributions of the best-fit parameters are well constrained and are shown in Fig.~\ref{corner_plot_all}. Also, the retrieved noise scaling terms have values close to one, indicating an appropriate error estimation. 

\begin{figure}
        \centering
        \includegraphics[width=0.5\textwidth]{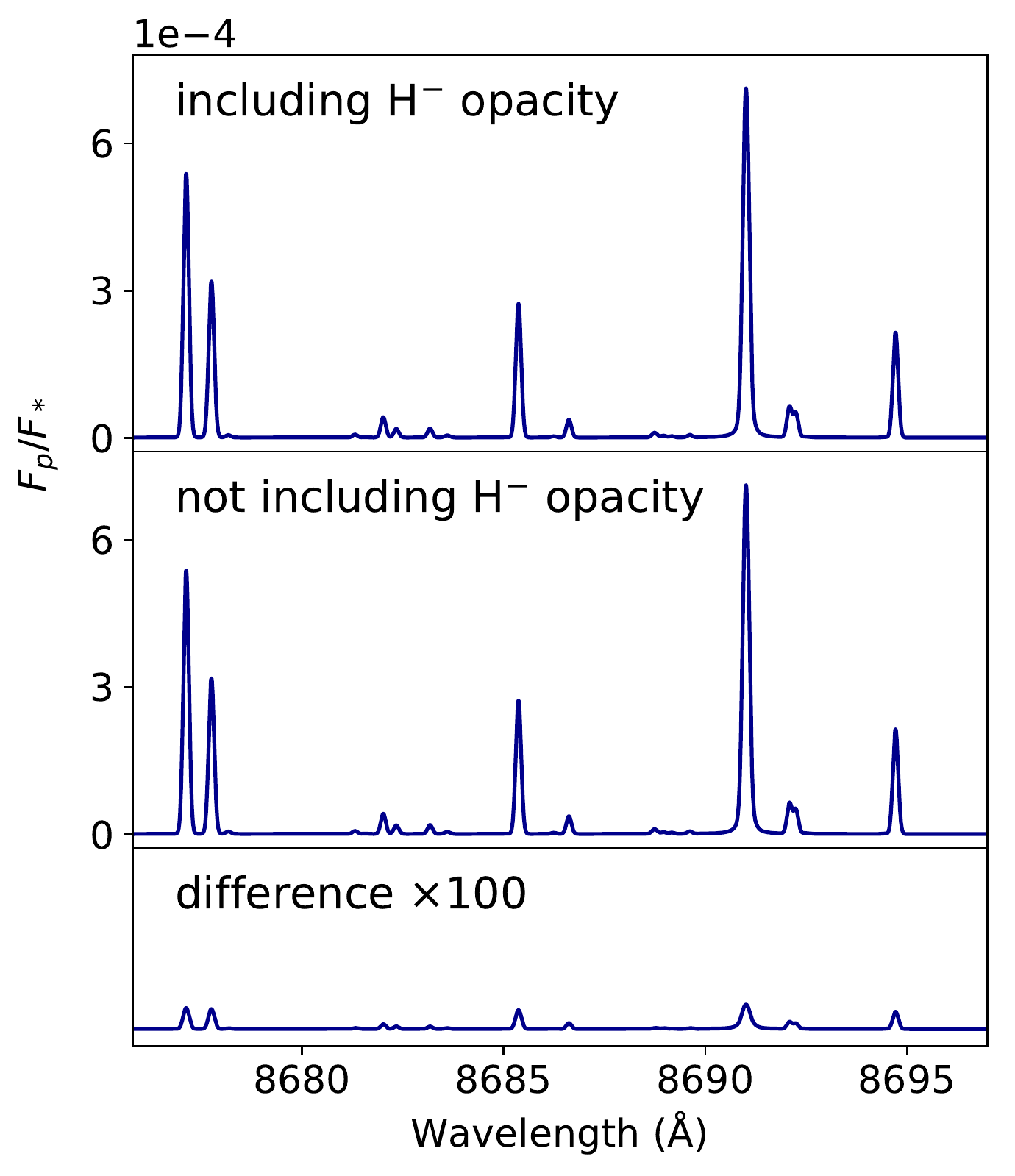}
        \caption{Comparison between model spectra including (\textit{top panel}) and not including (\textit{middle panel}) H$^-$ opacity. We used the emission lines of all the detected chemical species (i.e., \ion{Ti}{i}, \ion{V}{i}, OH, \ion{Fe}{i}, \ion{Si}{i}, and \ion{Ti}{ii}). The difference between the models is insignificant (\textit{bottom panel}). Therefore, including H$^-$ in the calculations would not affect our results significantly.}
        \label{Hminus_maps_comparison}
\end{figure}

Figure~\ref{retrieved-TP} illustrates the retrieved $T$-$p$ profile. A more exhaustive presentation is provided in Fig.~\ref{retrieved-TP_appendix_all}. We find that the inversion layer extends over the pressure range $10^{-5.1}$\,bar to $10^{-3.1}$\,bar, with temperatures of $3981_{-108}^{+213}$\,K and $3424_{-111}^{+107}$\,K in the upper and lower planetary atmosphere, respectively. The thermal inversion layer is weaker when compared to the retrieval results of similar UHJs \citep{Yan2020, Yan2022}. This implies that the emission lines of WASP-33b are relatively shallow. We investigated whether or not the presence of H$^-$ opacity could explain the relatively low intensity of the emission lines, but found that the species has a negligible impact on our results (Fig.~\ref{Hminus_maps_comparison}). To our knowledge, we performed the first retrieval of the thermal profile of WASP-33b's atmosphere using high-resolution spectroscopy. Similar work was recently carried out by \cite{vanSluijs2022}, who maximized the S/N detection strengths of the CO emission lines to study the atmosphere of the same planet. We retrieve temperatures that are consistent with their results at low atmospheric pressures, but are able to get tighter constraints on the thermal profile at higher pressures.

The retrieval is mainly driven by the spectral signature of the neutral chemical species, which we assume are mostly ionized in the upper atmospheric layers due to the elevated temperatures. Nevertheless, we suggest that the emission lines of neutral species emerging from these layers are strong enough to allow the determination of the temperature $T_1$ at the upper limit of the thermal inversion. This suggestion is motivated by the fact that the posterior distribution of $T_1$ is well constrained in Fig.~\ref{corner_plot_all}. In contrast, undetectable emission lines would cause a flat pattern toward higher temperatures in the posterior distribution, which is inconsistent with our results.

Our retrieval constrains the atmospheric metallicity to [M/H]\,=\,$1.49_{-0.76}^{+0.82}$\,dex, which corresponds to a supersolar elemental abundance in the upper planetary atmosphere. This is in line with the results of \cite{vanSluijs2022}, who obtain their strongest CO detection at abundances of $\sim$\,1\,dex in the atmosphere of WASP-33b. Moreover, our result is consistent with previous studies that measured atmospheric abundances greater than solar for a number of hot giant exoplanets \citep[e.g.,][]{Madhusudhan2014, Sedaghati2017, Nikolov2018}. 
A correlation of the inferred [M/H] with the pressure parameters $p_1$ and $p_2$ can be recognized as diagonal distribution patterns in Fig.~\ref{corner_plot_all}. This is consistent with the expected degeneracy between [M/H] and the atmospheric temperature profile. The VMRs of the investigated species are computed by assuming that the elemental abundances all vary with the same metallicity value. The reason for this approximation is that spectra with very high S/N would be needed to successfully determine the abundances of individual species, a task that may be addressed in future studies.

We point out that our forward model approximates the VMRs of the different species by assuming equilibrium chemistry. \cite{Fossati2021} on the other hand suggested that nonlocal thermodynamic equilibrium (NLTE) may play an important role in the upper atmospheres of UHJs. NLTE is expected to alter the population levels and thus the VMRs of different chemical species. For example, NLTE is predicted to strongly affect the population levels of Fe by lowering the VMR of \ion{Fe}{i} in favor of \ion{Fe}{ii}. The alteration of the population levels has important consequences for the amplitude of the planetary emission spectrum, the strength of atmospheric absorption of incoming stellar light, and therefore on the atmospheric $T$-$p$ profile. Given the great importance of Fe in the thermal inversion layers of UHJ atmospheres, future consideration of NLTE effects will increase the reliability of atmospheric retrievals.

We find that the spectral emission lines are significantly broadened. The retrieved Gaussian broadening profile has a standard deviation of $\varv_\mathrm{broad}$\,=\,$1.9\pm0.3$\,km\,s$^{-1}$, corresponding to a FWHM of $4.5\pm0.7$\,km\,s$^{-1}$. The thermal and pressure broadening information is already included in the forward model via the opacities of the radiative transfer calculation. Therefore, $\varv_\mathrm{broad}$ is expected to account only for the broadening effects from atmospheric dynamics and the rotation of the planet. In particular, rotational broadening is supposed to affect the spectral line width, given the high rotational velocity of $\sim$\,7\,km\,s$^{-1}$ at WASP-33b's equator when assuming tidal locking. The line broadening from our retrieval is lower than that expected from a planetary sphere where the flux is emitted homogeneously over its entire surface, which would yield a FWHM of $\sim$\,9\,km\,s$^{-1}$. We calculated this rotational broadening value with the \texttt{PyAstronomy} software package \citep{Czesla2019}. The retrieved line broadening therefore suggests that most of the contribution to WASP-33b’s emission spectrum may originate from the hottest region of the planetary atmosphere, which is located close to the substellar point.

We tested whether our method is capable of detecting the presence of spectral line broadening. To this end, we applied our retrieval method to synthetic data. We first took the model spectrum without additional broadening as shown in Fig.~\ref{SN_model_all-species}. We then broadened the spectrum with $\varv_\mathrm{broad}$\,=\,2\,km\,s$^{-1}$. Sequentially, we injected random white noise with different uncertainty levels to the model spectrum and performed the retrieval. We found that the correct broadening velocity can be retrieved even with the noise level of ten times the uncertainties of our observations (Fig.~\ref{broadening_test}). We therefore conclude that the quality of the observational data used is good enough to allow an appropriate determination of the spectral line broadening with our retrieval method.

\begin{table*}
        \caption{Atmospheric retrieval results on WASP-33b.
        }             
        \label{tab-retrieval-results}      
        \centering                          
        \renewcommand{\arraystretch}{1.4} 
        \begin{threeparttable}
                \begin{tabular}{l l l l l l l l l}        
                        \hline\hline                 
                        \noalign{\smallskip}
                        Parameter & All spectra & Close to eclipse & Far from eclipse & \ion{Fe}{i} & \ion{Ti}{i}~+~\ion{Ti}{ii} & \ion{V}{i}  & Unit  \\     
                        \noalign{\smallskip}
                        \hline                       
                        \noalign{\smallskip}
                        $T_1$                 & $3981_{-108}^{+213}$      & $4157_{-162}^{+274}$            & $3682_{-142}^{+302}$           & $3985_{-131}^{+167}$      & $4386_{-278}^{+333}$      & $4985_{-722}^{+714}$ &   K \\ 
                        $\log{p_1}$            & $-5.12_{-1.06}^{+0.93}$   & $-5.66_{-0.93}^{+1.10}$         & $-4.54_{-1.38}^{+1.03}$        & $-4.62_{-0.69}^{+0.36}$   & $-6.41_{-0.43}^{+0.69}$   & $-5.26_{-0.82}^{+0.58}$  &   log bar\\ 
                        $T_2$                 & $3424_{-111}^{+107}$      & $3525_{-131}^{+131}$             & $2831_{-997}^{+415}$          & $3459_{-203}^{+205}$      & $3497_{-315}^{+255}$      & $1949_{-648}^{+972}$ &   K \\ 
                        $\log{p_2}$            & $-3.08_{-0.54}^{+0.67}$   & $-3.05_{-0.63}^{+0.94}$         & $-2.71_{-0.77}^{+0.67}$        & $-3.31_{-0.33}^{+0.35}$   & $-2.93_{-0.39}^{+0.35}$   & $-3.02_{-0.51}^{+0.95}$   & log bar\\ 
                        $\mathrm{[M/H]}$      & $1.49_{-0.76}^{+0.83}$    & $1.46_{-1.01}^{+1.13}$               & $1.23_{-0.81}^{+0.98}$                   & 1.49 (fixed)              & 1.49 (fixed)              & 1.49 (fixed) &   dex \\ 
                        $\varv_\mathrm{broad}$       & $1.9\pm0.3$    & $2.1_{-0.4}^{+0.3}$          & $1.7_{-0.6}^{+0.5}$         & $2.4_{-0.4}^{+0.3}$    & $1.6\pm0.8$    & $1.7_{-0.9}^{+1.0}$   & km\,s$^{-1}$\\ 
                        \noalign{\smallskip}
                        \hline                       
                        \noalign{\smallskip}                    
                        $\beta_\mathrm{CV}$   & $0.841\pm0.002$           & $0.842\pm0.002$                  & $0.842\pm0.002$                 & $0.842\pm0.002$           & $0.842\pm0.002$           & $0.844\pm0.002$   & \ldots\\ 
                        $\beta_\mathrm{CN}$   & $0.641\pm0.002$           & $0.639\pm0.002$                  & $0.639_{-0.001}^{+0.002}$               & $0.641\pm0.002$           & $0.641\pm0.002$           & $0.640\pm0.002$   & \ldots\\ 
                        $\beta_\mathrm{H}$    & $0.795\pm0.001$           & $0.800\pm0.001$                 & $0.789\pm0.001$                & $0.795\pm0.001$           & $0.795\pm0.001$           & $0.798\pm0.001$   & \ldots\\ 
                        $\beta_\mathrm{E}$    & $1.122\pm0.002$           & $1.101\pm0.002$                 & $1.155\pm0.002$                & $1.122\pm0.002$           & $1.122\pm0.002$           & $1.124\pm0.002$   & \ldots\\ 
                        \noalign{\smallskip}
                        \hline                                   
                \end{tabular}
                \tablefoot{In total, we ran six different retrievals. Three retrievals were performed with the spectral lines of all the detected species (i.e., \ion{Ti}{i}, \ion{V}{i}, OH, \ion{Fe}{i}, \ion{Si}{i}, and \ion{Ti}{ii}). We used (i) all the spectra, (ii) a subset of spectra at orbital phases close to the secondary eclipse, and (iii) a subset of spectra far from the secondary eclipse. Another three retrievals were performed by including all the spectra, but only with the emission lines of (iv) \ion{Fe}{i}, (v) \ion{Ti}{i}~+~\ion{Ti}{ii} combined, and (vi) \ion{V}{i}, respectively. We conducted the retrievals by using uniform priors with the boundaries as follows. $T_1$ and $T_2$: 1000 to 6000\,K; $p_1$ and $p_2$: $10^{-7}$ to 1\,bar; [M/H]: --3 to 3\,dex; $\varv_\mathrm{broad}$: 0 to 10\,km\,s$^{-1}$; $\beta_\mathrm{CV}$, $\beta_\mathrm{CN}$, $\beta_\mathrm{H}$, and $\beta_\mathrm{E}$: 0 to 3.}
        \end{threeparttable}      
\end{table*}

\subsubsection{Phase resolved retrieval}
\label{Phase resolved retrieval}

The observer's line of sight aligns with different geographical longitudes of the planet as a function of the orbital phase. Consequently, different regions of the planetary atmosphere are expected to contribute to the observed signal during our observations. Performing the retrieval on subsets of the spectral time series that correspond to different orbital phase intervals will allow us to study the physical conditions at different longitudes in the planetary atmosphere.  

To perform a phase-resolved retrieval, we subdivided the spectral time series into two subsets for calculating the master residual spectra. One subset corresponds to the spectra close to the secondary eclipse when the dayside hemisphere faces the observer. The other subset consists of the spectra far from the secondary eclipse, and therefore contains the information from regions of both the planetary day- and nightsides. The orbital phase coverage of the two subsets is illustrated in Fig.~\ref{retrieved-TP}. To avoid any interference with residual stellar lines, only the spectra at orbital phases outside the RV interval $\pm \varv_\mathrm{rot}\sin{i_*}$ were used.

The posterior distributions of the fit parameters are shown in Figs.~\ref{corner_plot_dayside} and \ref{corner_plot_terminator}. Figure~\ref{retrieved-TP} compares the $T$-$p$ curves obtained from the two subsets with the thermal profile obtained from all the spectra. A more detailed overview on the $T$-$p$ profiles is also shown in Fig.~\ref{retrieved-TP_appendix_all}. Depending on the pressure level, the thermal profiles calculated from the two spectral subsets differ by about 300\,K to 700\,K, and the profile determined from all the spectra lies between the two. We find that the spectra close to the eclipse deliver a hotter temperature profile in comparison to those far from the eclipse. As the retrieved temperature profile corresponds to the average $T$-$p$ curve of the visible hemisphere, this is consistent with the expectation of a temperature gradient away from the substellar point. Close to the secondary eclipse, mainly the dayside is facing the observer, leading to a hot average temperature profile. However, for spectra far from the secondary eclipse, the terminator region and a significant fraction of the cooler nightside are aligned with the observer's line of sight, resulting in lower atmospheric temperatures being measured.

The values of [M/H] and $\varv_\mathrm{broad}$ obtained for the two spectral subsets are consistent. We note a trend that spectra far from the secondary eclipse are less affected by line broadening than spectra near the eclipse. This is a reasonable result because far from the secondary eclipse, a significant fraction of the planetary disk that faces the observer is not illuminated by the host star. The nonilluminated planetary atmosphere is not expected to significantly contribute to the emission signal, decreasing the impact of the rotational line broadening. The dependence of the line broadening from the orbital phase can also be recognized when considering the CCF trail in Fig.~\ref{SN_model_all-species}, showing an increased width of the CCF toward the secondary eclipse. However, we note that the retrieved $\varv_\mathrm{broad}$ values should be compared with caution because of the relatively large uncertainties. All results of the phase resolved retrieval are listed in Table~\ref{tab-retrieval-results}.

\subsubsection{Retrieval of individual species}

\begin{figure}
        \centering
        \includegraphics[width=0.5\textwidth]{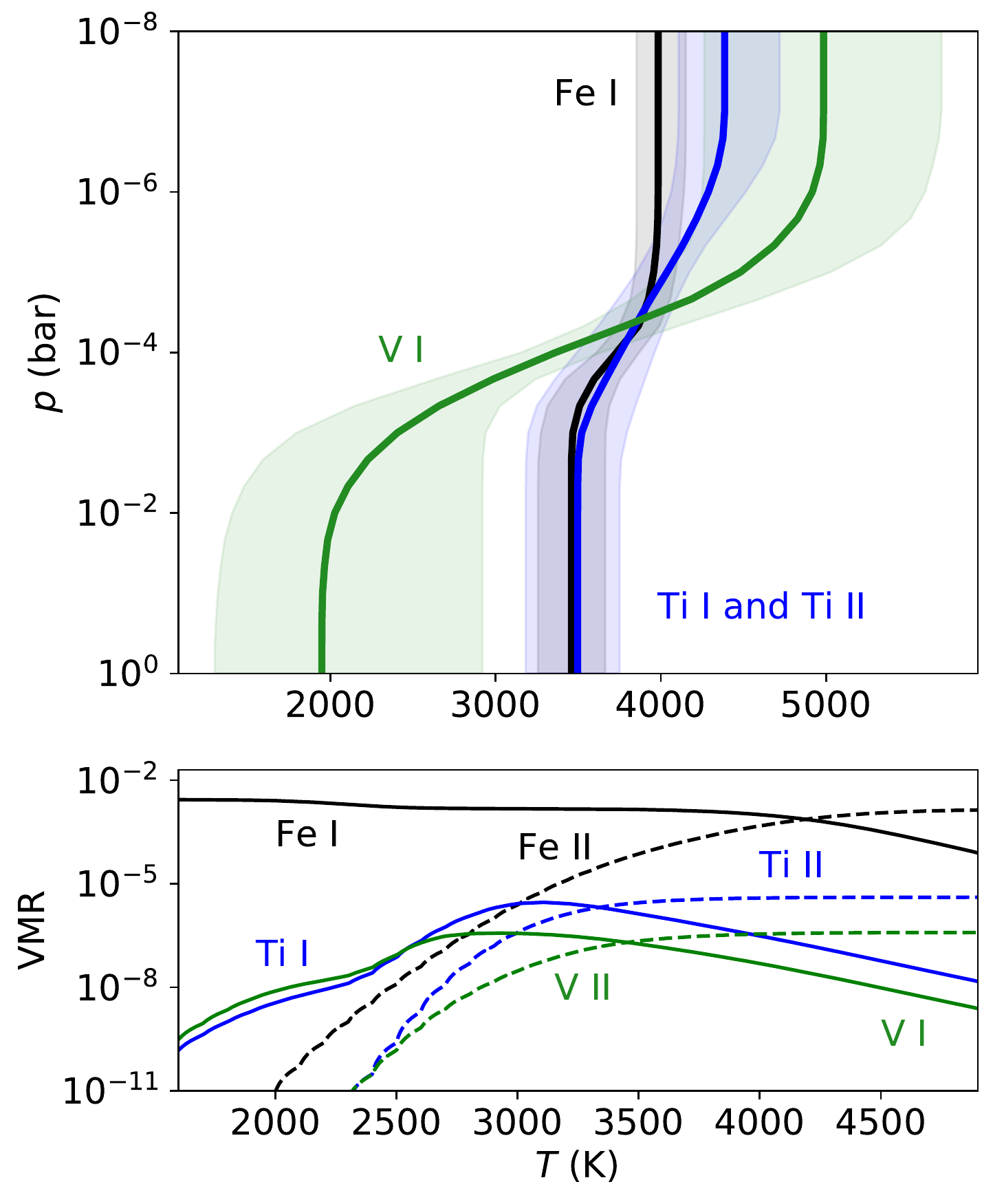}
        \caption{$T$-$p$ profiles retrieved with the lines from individual chemical species and VMRs as a function of temperature. The temperature profiles retrieved from the \ion{Ti}{i} and \ion{Ti}{ii}, \ion{V}{i}, and \ion{Fe}{i} emission lines are shown in the \textit{top panel}. We report the median of the sampled temperature profiles and the 1$\sigma$ interval. The \textit{bottom panel} shows the chemical equilibrium VMRs of \ion{Ti}{i}, \ion{Ti}{ii}, \ion{V}{i}, \ion{V}{ii}, \ion{Fe}{i}, and \ion{Fe}{ii} as a function of temperature. We set [M/H] to the retrieved value of 1.49\,dex; the pressure is set to $10^{-4}$\,bar. Ti and V are expected to undergo stronger ionization than Fe. 
        }
        \label{retrieved-TP_Fe-Ti-V}
\end{figure}

To compare the temperature profiles probed by individual chemical species, we ran retrievals for Ti, V, and Fe. These are the species with the most prominent detections in Sect.~\ref{Cross-correlation method}. Retrieving the $T$-$p$ profile for the individual species also allows us to test the consistency of the results of our method. We set [M/H] to the previously determined value of 1.49\,dex (cf. Sect.~\ref{Retrieval including all the detected chemical species}), because the metallicity was poorly constrained when leaving it as a free parameter. For the Ti retrieval, we included the opacities of both \ion{Ti}{i} and \ion{Ti}{ii}. As we were not able to constrain the presence of \ion{V}{ii} and \ion{Fe}{ii} by cross-correlation, only \ion{V}{i} and \ion{Fe}{i} were included in the other two retrievals, respectively. We used the spectra with a planetary RV outside $\pm \varv_\mathrm{rot}\sin{i_*}$ for the retrievals including \ion{Ti}{ii} and \ion{Fe}{i}, which are affected by residual stellar lines. For \ion{V}{i}, we included all the out-of-eclipse spectra because we did not find any significant residuals of stellar \ion{V}{i} lines. The posterior distributions of the fit parameters are shown in Figs.~\ref{corner_plot_Fe}, \ref{corner_plot_Ti-TiII}, and \ref{corner_plot_V}. Figure~\ref{retrieved-TP_Fe-Ti-V} compares the retrieved $T$-$p$ profiles and a detailed overview is provided in Fig.~\ref{retrieved-TP_appendix_Fe-Ti-V}. All results are summarized in Table~\ref{tab-retrieval-results}.

For \ion{Fe}{i}, the parameters are well constrained and almost perfectly match the results from the retrieval that uses all the species in Sect.~\ref{Retrieval including all the detected chemical species}. We attribute this similarity to the larger number of \ion{Fe}{i} lines when compared to that of all the other species. This also suggests that the retrieval that includes all the species is mainly driven by the \ion{Fe}{i} lines. Also, the $T$-$p$ profile from \ion{Ti}{i} and \ion{Ti}{ii} is in agreement with the retrieval result of all the species. The retrieval yielded a well constrained set of fit parameters, albeit at a somewhat lower precision than for \ion{Fe}{i}. This is probably caused by the lower number and decreased strength of the \ion{Ti}{i} and \ion{Ti}{ii} spectral lines in comparison to the \ion{Fe}{i} emission signature. The fact that retrievals for independent species achieve consistent results demonstrates the reliability of our method and gives confidence to the calculated parameter values.

The posterior distributions of the \ion{V}{i} retrieval provide a significantly poorer constraint of the atmospheric parameters. This results in extended uncertainties of the $T$-$p$ profile in the lower and upper planetary atmosphere. In view of the relatively low detection significance and the scarce number of features in the \ion{V}{i} model spectrum in Fig.~\ref{SNmaps_and_template_spectra_neutral}, this agrees with our expectations. While the temperature profile inferred from \ion{V}{i} is close to that derived from \ion{Ti}{i}/\ion{Ti}{ii} and \ion{Fe}{i} in the upper planetary atmosphere, we find a deviation at pressures higher than $10^{-3}$\,bar. Apart from the large uncertainties, the interpretation of this difference is unfortunately difficult, because both physical and method-dependent effects can cause the observed discrepancy. One possible scenario could be that the individual species are confined to different regions in the planetary atmosphere, and each has its own temperature profile. Figure~\ref{retrieved-TP_Fe-Ti-V} shows the chemical equilibrium VMRs of \ion{Ti}{i}, \ion{Ti}{ii}, \ion{V}{i}, \ion{V}{ii}, \ion{Fe}{i}, and \ion{Fe}{ii} as a function of temperature. We consider a pressure level of $10^{-4}$\,bar, which corresponds to the location of the thermal inversion layer. The VMRs of \ion{Ti}{i} and \ion{Ti}{ii} show a similar pattern when compared to that of \ion{V}{i} and \ion{V}{ii}, respectively. The colder temperatures probed by \ion{V}{i} at higher pressure levels could therefore be explained by a lower abundance of the species than assumed by our method. In this case, the hottest region of the planetary atmosphere would be depleted by \ion{V}{i} due to ionization and the hypothesized low abundance. Therefore, the \ion{V}{i} signature would encode mainly the information from the coolest regions of the planetary dayside, leading to a more moderate $T$-$p$ profile. The nominal abundance of \ion{Ti}{i} and the inclusion of \ion{Ti}{ii} would instead allow information from the entire dayside atmosphere to be considered, resulting in a hotter $T$-$p$ profile. Alternatively, the use of oversimplified modeling of the VMRs could prevent an accurate estimation of the temperature profile for individual species. If the equilibrium chemistry assumption is not met, this could result in a poorly constrained temperature curve due to the degeneracy between temperature and VMR. Also, an oversimplification of the $T$-$p$ model used and the presence of horizontal variations of the temperature profile from the substellar point to the nightside could explain the discrepancy measured. Spectra with increased S/N, the inclusion of nonequilibrium chemistry, and the use of a more comprehensive $T$-$p$ model will enable us to better understand the information encoded in the spectral signature of individual species.

\section{Conclusions}
\label{Conclusions}

We used observations from the high-resolution spectrographs CARMENES, HARPS-N, and ESPaDOnS to analyze the thermal emission spectrum of the UHJ WASP-33b. A joint analysis of the data from the three instruments allowed us to investigate the planetary emission spectrum over an extended wavelength range from the near-ultraviolet to the NIR (3700--17\,100\,\AA). We applied the cross-correlation technique to extract the faint spectral signature of the planetary atmosphere from the observations. This analysis led to the first detection of the emission signature of \ion{Ti}{i} and \ion{V}{i} in an exoplanet atmosphere, as these two species had previously been detected exclusively by transmission spectroscopy. Also, we detected a tentative emission signal of \ion{Ti}{ii}. These detections are an important finding, given the frequently observed depletion of Ti- and V-bearing species in the atmospheres of UHJs. Moreover, we confirmed the presence of OH, \ion{Fe}{i}, and \ion{Si}{i} detected by previous studies. No significant signature from the ionic species \ion{V}{ii}, \ion{Fe}{ii}, or \ion{Si}{ii} could be found in our spectral time series. The identification of spectral emission lines unambiguously proves the presence of a thermal inversion layer in the dayside atmosphere of WASP-33b, which is in line with theoretical work on highly irradiated planetary atmospheres.

We conducted a retrieval for the atmospheric $T$-$p$ profile, the elemental abundances, and the spectral line broadening. The retrieval was performed using the data from all three instruments together. For this purpose, we forward modeled the emission lines of all the detected species via the radiative transfer code \texttt{petitRADTRANS} and assumed equilibrium chemistry. Compared to other UHJs (e.g., WASP-189b, KELT-20b/MASCARA-2b), our retrieval results in a relatively weak thermal inversion that extends roughly from 3400\,K to 4000\,K at pressures near $10^{-3}$\,bar and $10^{-5}$\,bar. We determined supersolar elemental abundances around 1.5\,dex in the upper planetary atmosphere and found a spectral line broadening with a Gaussian FWHM of about 4.5\,km\,s$^{-1}$. By running the retrieval on two distinct subsets of the spectral time series, we obtained temperature profiles that differ by 300\,K to 700\,K. This confirms the expectation that the temperature is higher on the planetary dayside than on the nightside. We also performed the retrieval for different chemical species individually. The temperature profiles from \ion{Ti}{i}/\ion{Ti}{ii} and \ion{Fe}{i} are in good agreement with the overall result from all the species. However, the $T$-$p$ profile of the \ion{V}{i} signature slightly deviates from that of the other species. We suggest that a \ion{V}{i}-depleted planetary atmosphere could explain the measured discrepancy between the $T$-$p$ profiles.

Our work shows that high-resolution emission spectroscopy offers the possibility to study the physical conditions in UHJ atmospheres in great detail. Further progress on atmospheric retrievals will be achieved by deploying more sophisticated spectral forward models, increasing the number of species included, and expanding the orbital phase coverage of UHJ observations.

%

\begin{acknowledgements}

    The authors thank the referee for very useful comments and suggestions.
        
        CARMENES is an instrument at the Centro Astron\'omico Hispano en Andaluc{\'i}a (CAHA) at Calar Alto (Almer\'{\i}a, Spain), operated jointly by the Junta de Andaluc\'ia and the Instituto de Astrof\'isica de Andaluc\'ia (CSIC).
        
        CARMENES was funded by the Max-Planck-Gesellschaft (MPG), 
        the Consejo Superior de Investigaciones Cient\'{\i}ficas (CSIC),
        the Ministerio de Econom\'ia y Competitividad (MINECO) and the European Regional Development Fund (ERDF) through projects FICTS-2011-02, ICTS-2017-07-CAHA-4, and CAHA16-CE-3978, 
        and the members of the CARMENES Consortium 
        (Max-Planck-Institut f\"ur Astronomie,
        Instituto de Astrof\'{\i}sica de Andaluc\'{\i}a,
        Landessternwarte K\"onigstuhl,
        Institut de Ci\`encies de l'Espai,
        Institut f\"ur Astrophysik G\"ottingen,
        Universidad Complutense de Madrid,
        Th\"uringer Landessternwarte Tautenburg,
        Instituto de Astrof\'{\i}sica de Canarias,
        Hamburger Sternwarte,
        Centro de Astrobiolog\'{\i}a and
        Centro Astron\'omico Hispano-Alem\'an), 
        with additional contributions by the MINECO, 
        the Deutsche Forschungsgemeinschaft (DFG) through the Major Research Instrumentation Programme and Research Unit FOR2544 ``Blue Planets around Red Stars'', 
        the Klaus Tschira Stiftung, 
        the states of Baden-W\"urttemberg and Niedersachsen, 
        and by the Junta de Andaluc\'{\i}a.
        
        We acknowledge financial support from 
        the DFG through Research Unit FOR2544 ``Blue Planets around Red Stars'' (RE 1664/21-1);
        the Agencia Estatal de Investigaci\'on of the Ministerio de Ciencia e Innovaci\'on and the ERDF ``A way of making Europe'' through projects 
PID2019-109522GB-C5[1:4], 
PID2019-110689RB-I00, 
        and the Centre of Excellence ``Severo Ochoa'' and ``Mar\'ia de Maeztu'' awards to the Instituto de Astrof\'isica de Canarias (CEX2019-000920-S), Instituto de Astrof\'isica de Andaluc\'ia (SEV-2017-0709), and Centro de Astrobiolog\'ia (MDM-2017-0737); 
the European Research Council under the European Union’s Horizon 2020 research and innovation program under grant agreements Nos. 
832428-Origins 
and 694513, 
and under the Marie Sk\l{}odowska-Curie grant agreement No. 895525;
and the Generalitat de Catalunya/CERCA programme.

\end{acknowledgements}

\bibliographystyle{aa} 

\bibliography{W33-retrieval-refer}

\appendix

        \FloatBarrier
        \clearpage

        \section{Posterior distributions}

        \begin{figure}[H]
                \onecolumn
                \centering
                \includegraphics[width=0.62\textwidth]{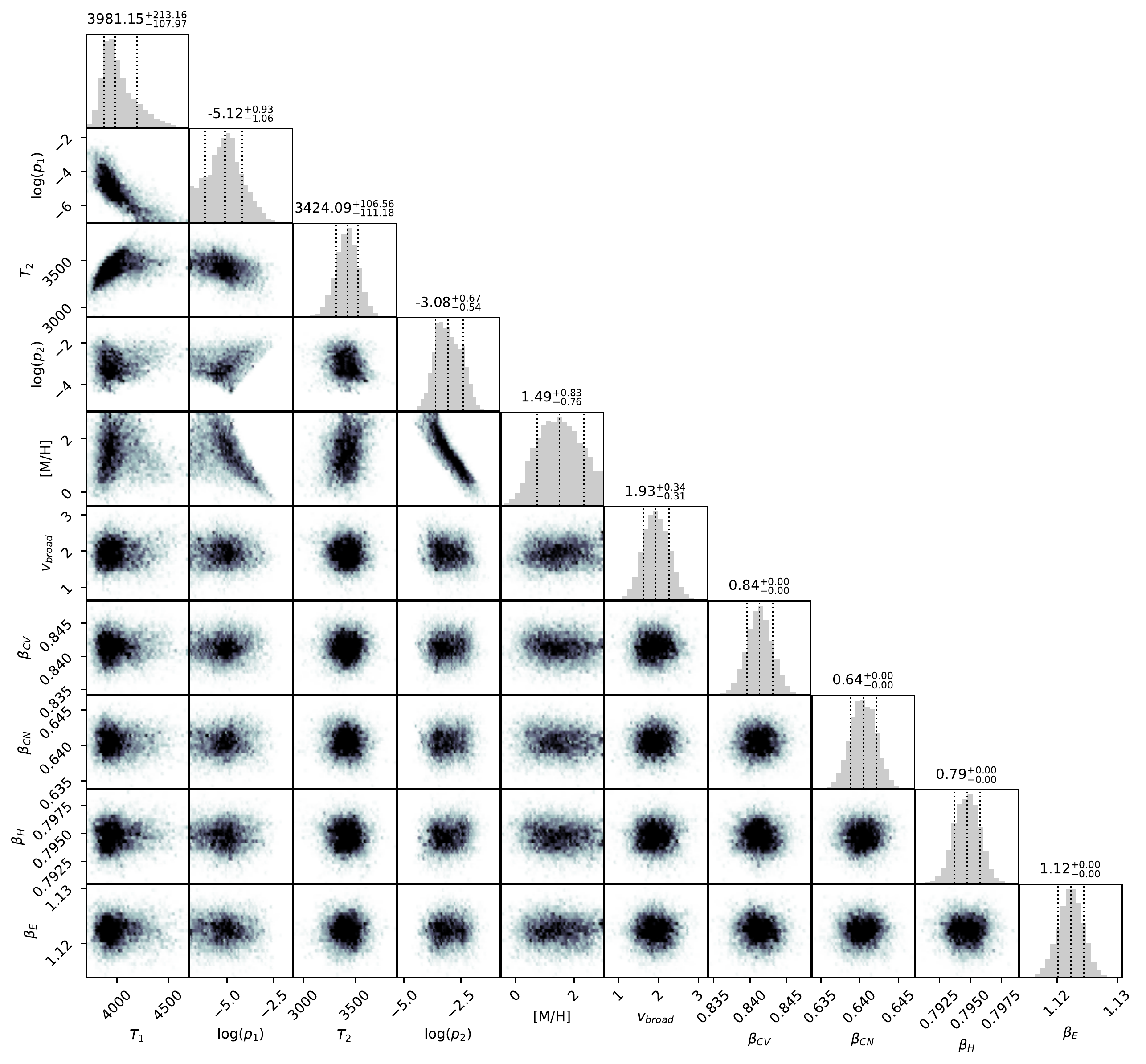}
                \caption{Posterior distributions of the atmospheric parameters from our retrieval. We used the emission lines of all the detected species (i.e., \ion{Ti}{i}, \ion{V}{i}, OH, \ion{Fe}{i}, \ion{Si}{i}, and \ion{Ti}{ii}).}
                \label{corner_plot_all}
        \end{figure}

        \begin{figure}[H]
        \onecolumn
                \centering
                \includegraphics[width=\textwidth]{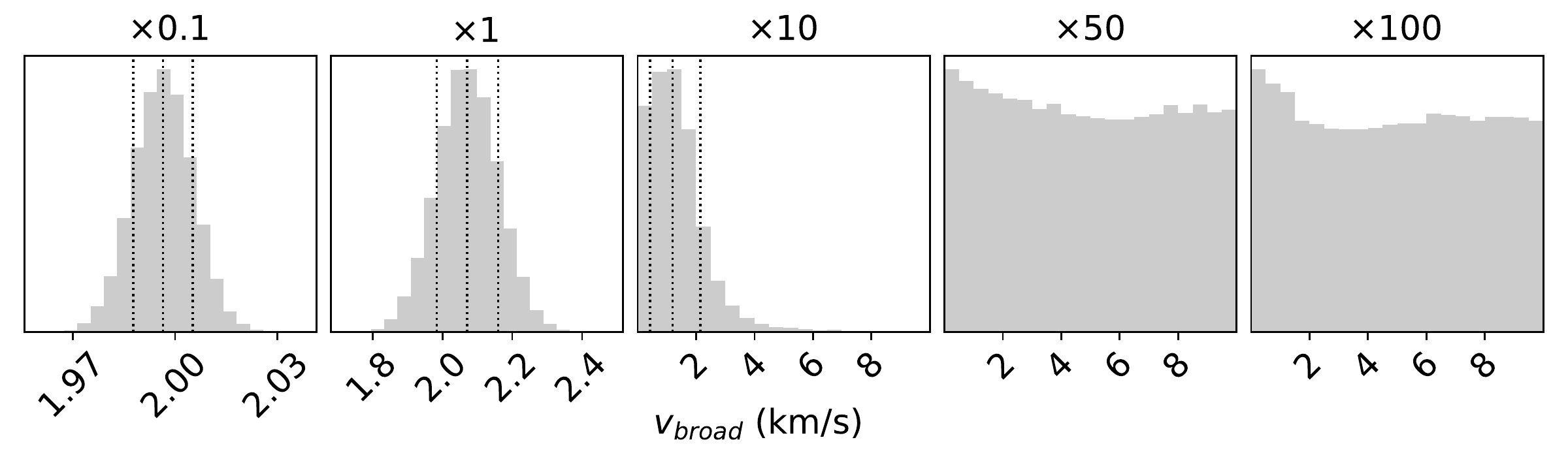}
                \caption{Posterior distributions of $\varv_\mathrm{broad}$ from simulated data with different noise levels. The noise level corresponds to that of the observed spectra multiplied by the factor that is reported at the top of each panel. The simulated broadening of 2\,km\,s$^{-1}$ can be retrieved for noise levels up to ten times that of the simulated observation data. The median and 1$\sigma$ percentiles of the distributions that retrieve the simulated line broadening are indicated by the dashed vertical lines.}
                \label{broadening_test}
        \end{figure}

        \begin{figure*}
                \centering
                \includegraphics[width=0.62\textwidth]{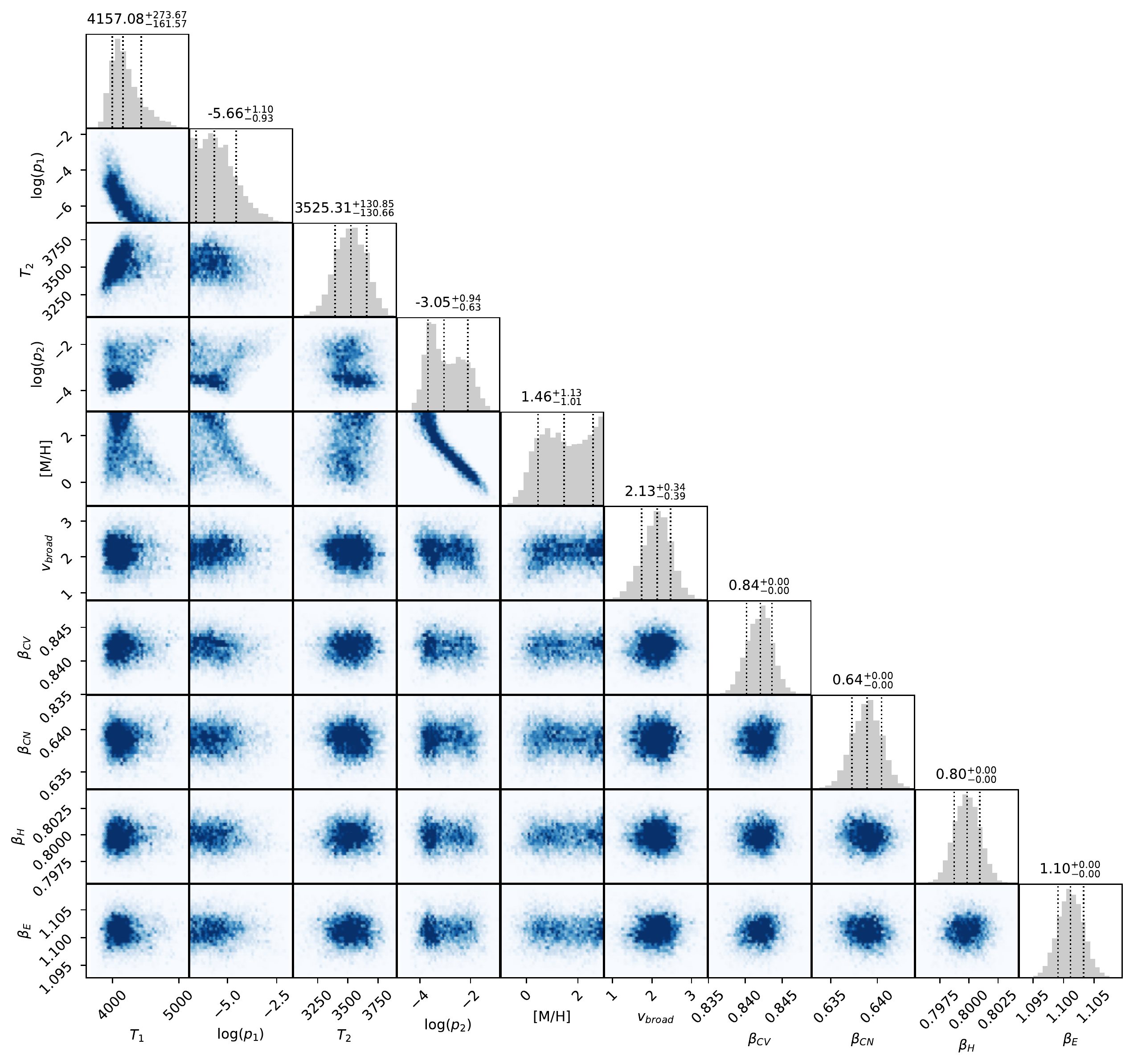}
                \caption{Same as Fig.~\ref{corner_plot_all}, but considering only the spectra close to the secondary eclipse.}
                \label{corner_plot_dayside}
        \end{figure*}

        \begin{figure*}
                \centering
                \includegraphics[width=0.62\textwidth]{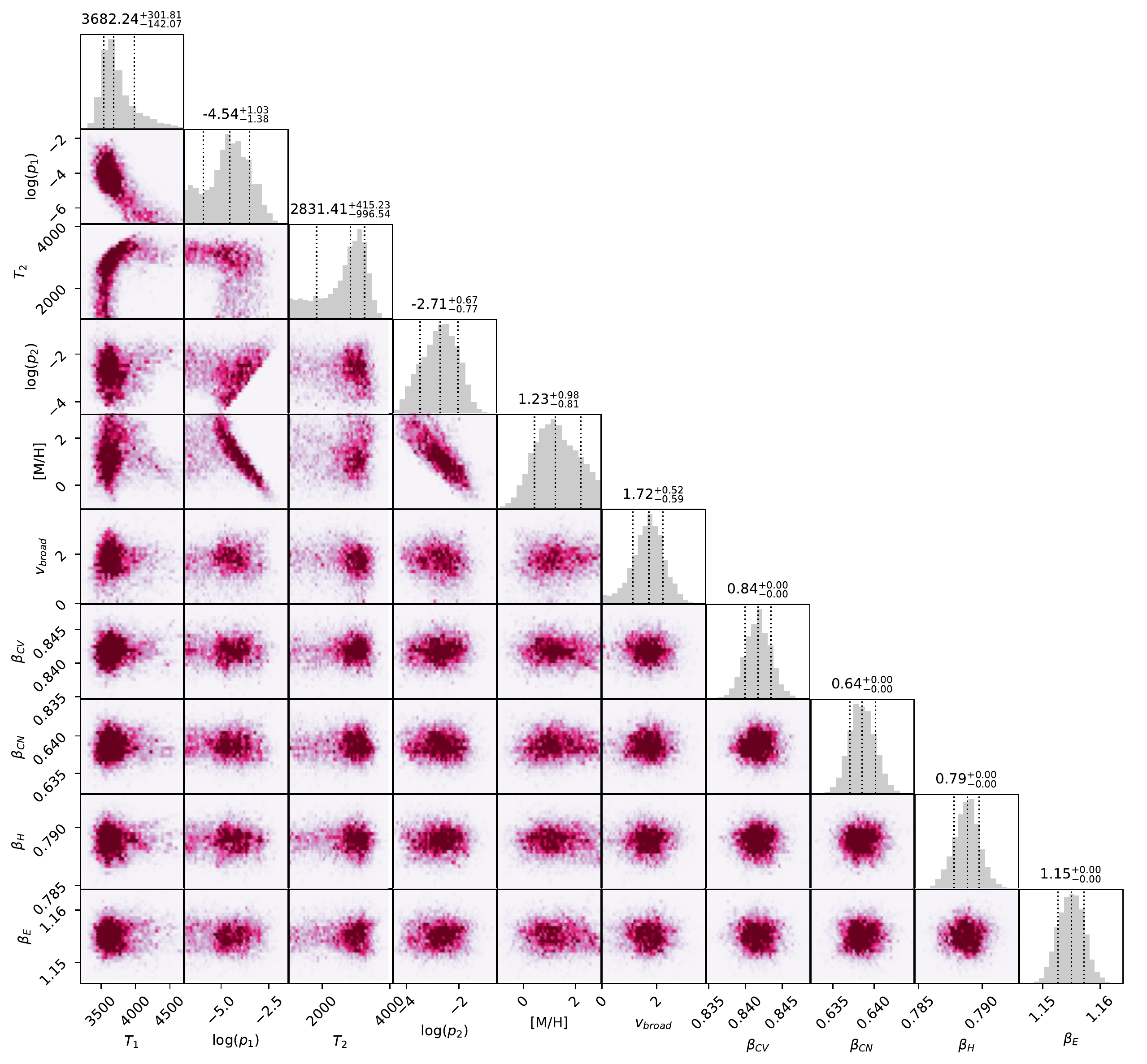}
                \caption{Same as Fig.~\ref{corner_plot_all}, but considering only the spectra far from the secondary eclipse.}
                \label{corner_plot_terminator}
        \end{figure*}
        
        \begin{figure*}
                \centering
                \includegraphics[width=0.62\textwidth]{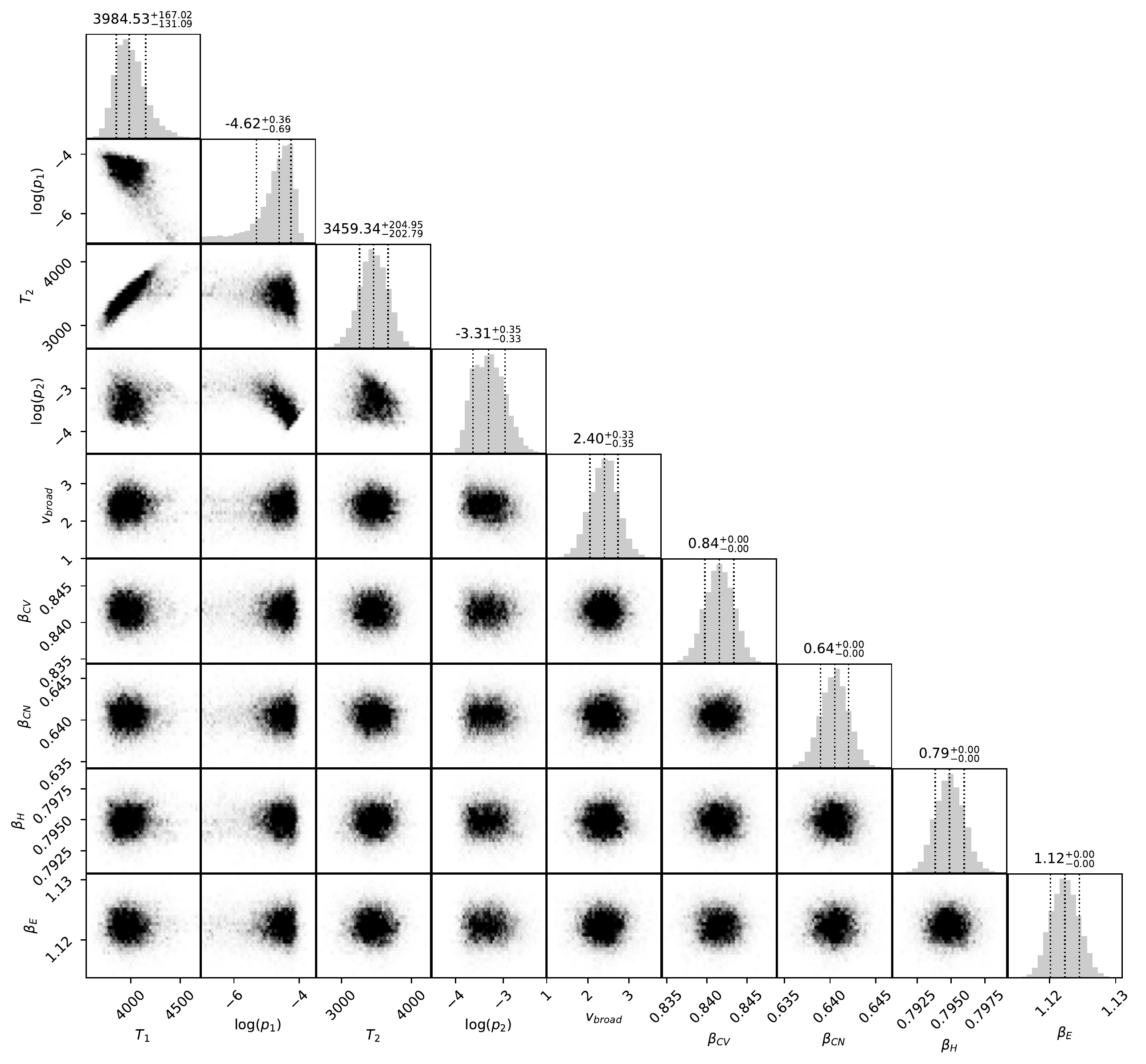}
                \caption{Same as Fig.~\ref{corner_plot_all}, but computed with the opacities from \ion{Fe}{i} only. The metallicity was fixed to [M/H]\,=\,1.49\,dex.}
                \label{corner_plot_Fe}
        \end{figure*}
        
        \begin{figure*}
                \centering
                \includegraphics[width=0.62\textwidth]{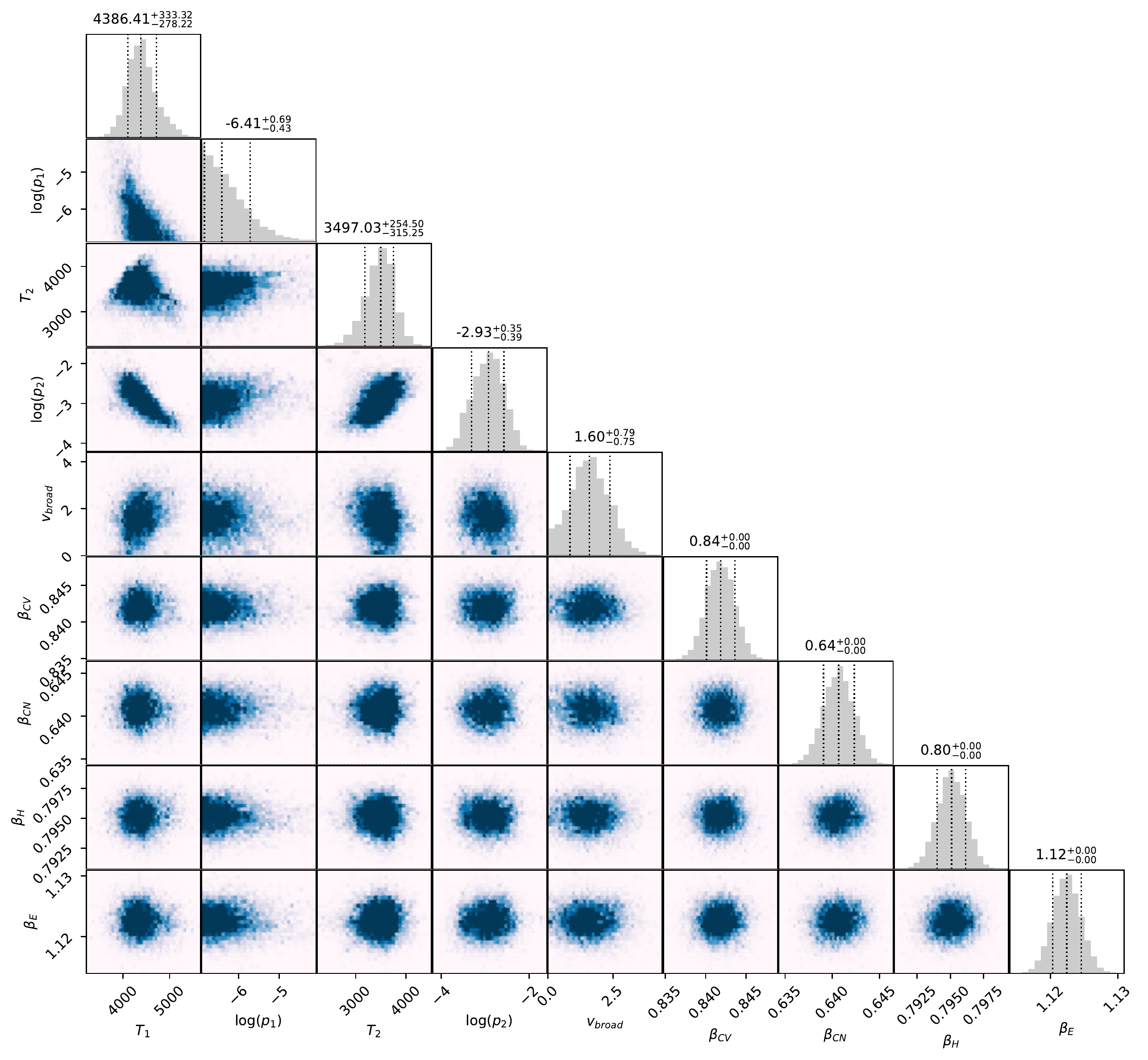}
                \caption{Same as Fig.~\ref{corner_plot_all}, but computed with the opacities from \ion{Ti}{i} and \ion{Ti}{ii} only. The metallicity was fixed to [M/H]\,=\,1.49\,dex.}
                \label{corner_plot_Ti-TiII}
        \end{figure*}
        
        \begin{figure*}
                \centering
                \includegraphics[width=0.62\textwidth]{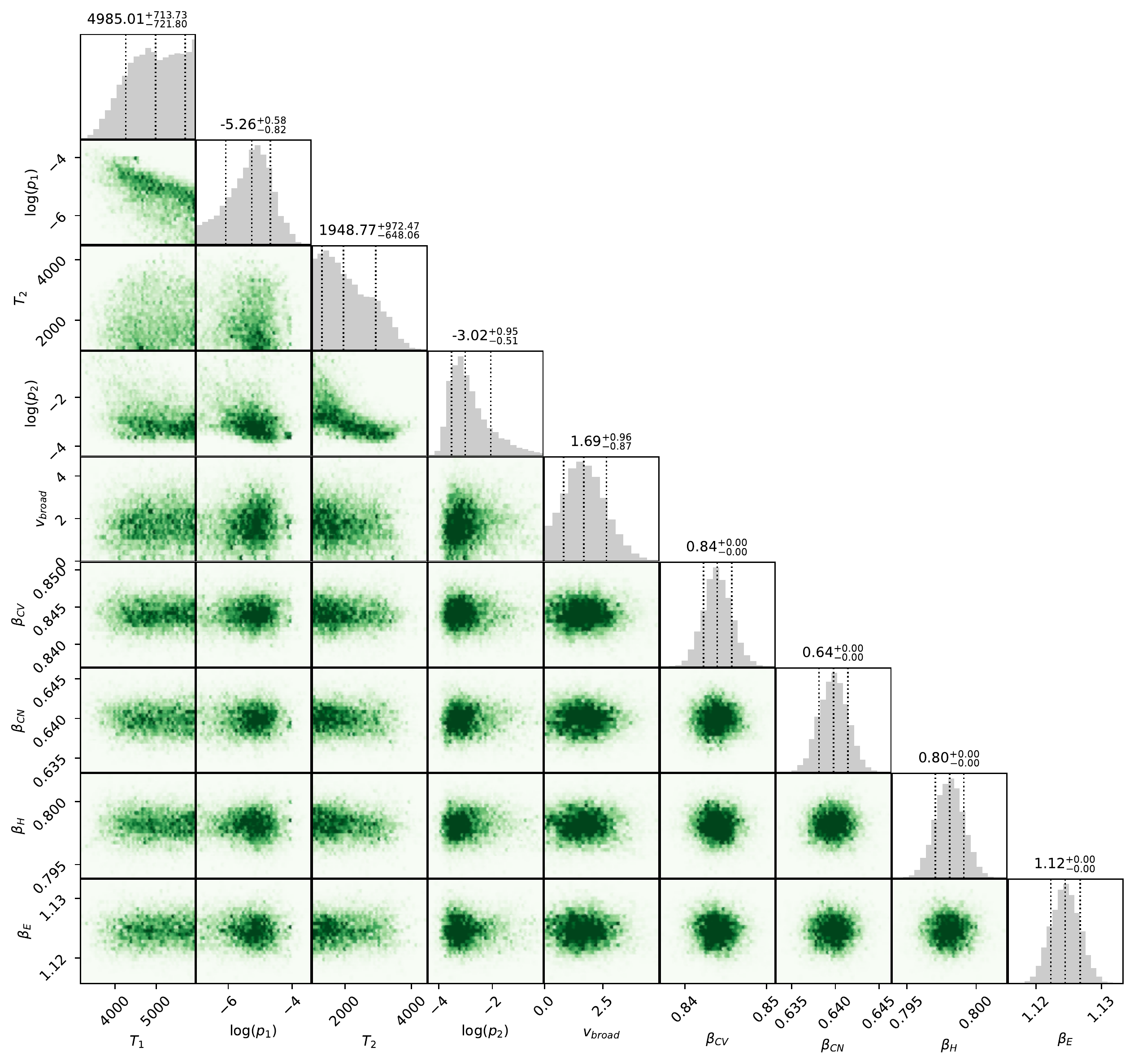}
                \caption{Same as Fig.~\ref{corner_plot_all}, but computed with the opacities from \ion{V}{i} only. The metallicity was fixed to [M/H]\,=\,1.49\,dex.}
                \label{corner_plot_V}
        \end{figure*}
        
        \FloatBarrier
        \clearpage
        
        \section{Thermal profiles}

        \begin{figure}[H]
        \onecolumn
                \centering
                \includegraphics[width=\textwidth]{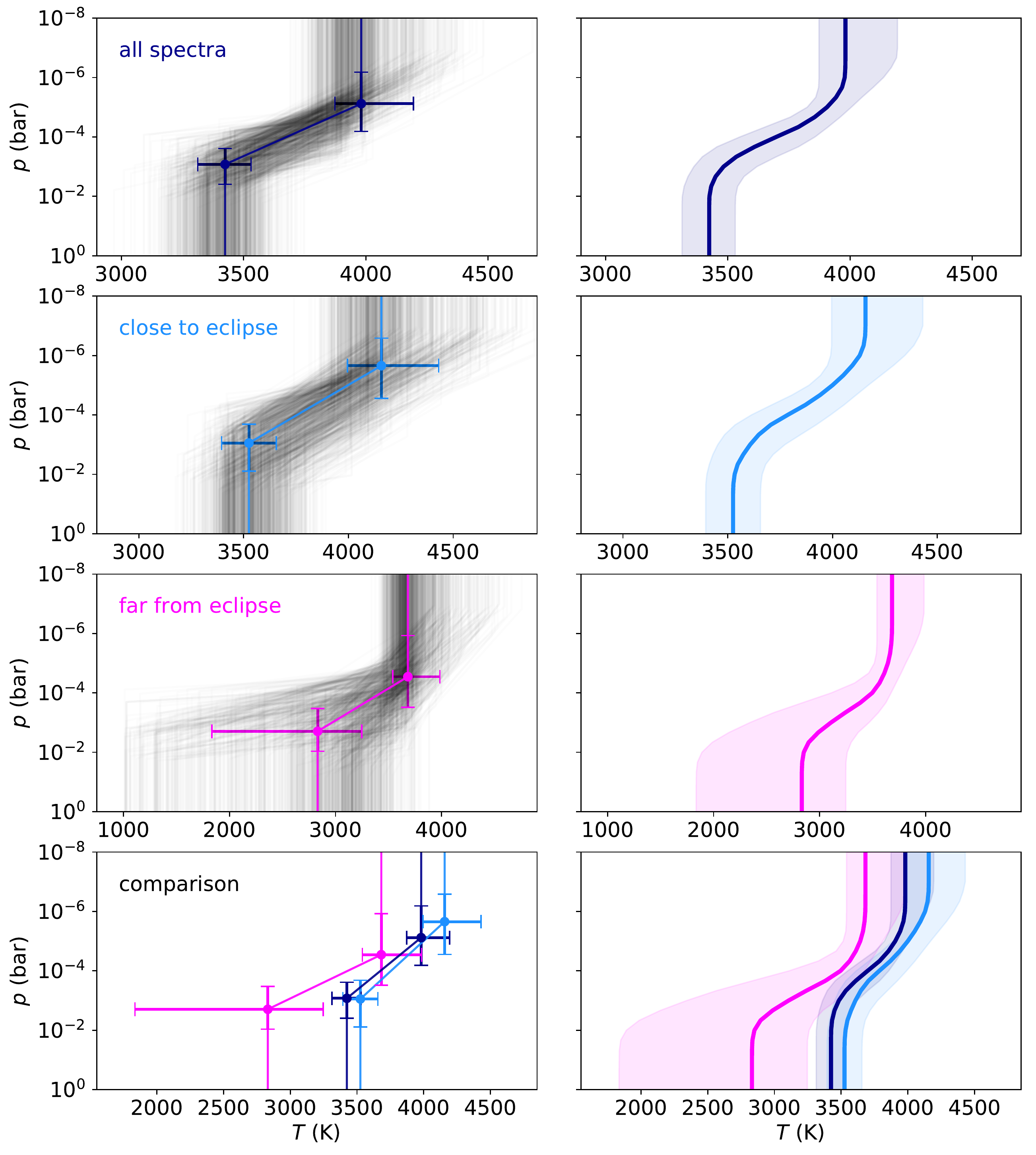}
                \caption{Retrieved $T$-$p$ profiles using the spectral lines of all the detected chemical species (i.e., \ion{Ti}{i}, \ion{V}{i}, OH, \ion{Fe}{i}, \ion{Si}{i}, and \ion{Ti}{ii}). The \textit{left panels} show the retrieved two-point $T$-$p$ profiles with the respective uncertainties. Examples of the $T$-$p$ profiles that were sampled when running the MCMC method are indicated in gray. The \textit{right panels} show the median of the sampled temperature profiles and the 1$\sigma$ interval. The \textit{panels} from top to bottom show the temperature profiles obtained using all the spectra, a subset of spectra close to the secondary eclipse, a subset of spectra far from the secondary eclipse, and a comparison of the three $T$-$p$ curves.}
                \label{retrieved-TP_appendix_all}
        \end{figure}

        \begin{figure*}
                \centering
                \includegraphics[width=\textwidth]{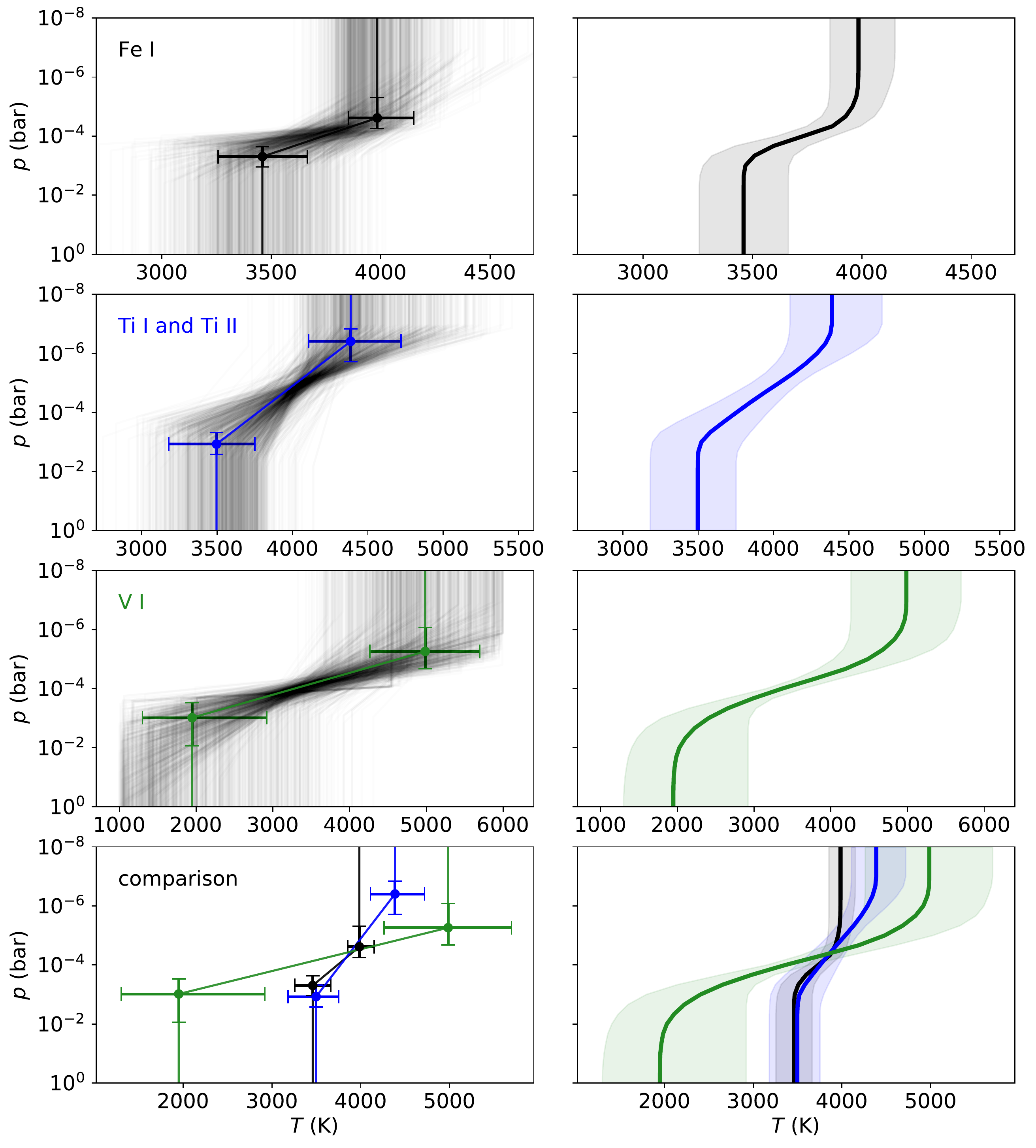}
                \caption{Same as Fig.~\ref{retrieved-TP_appendix_all}, but for individual chemical species. The \textit{panels} from top to bottom show the temperature profiles obtained with the emission lines of \ion{Fe}{i}, \ion{Ti}{i} and \ion{Ti}{ii}, \ion{V}{i}, and a comparison of the three $T$-$p$ curves.}
                \label{retrieved-TP_appendix_Fe-Ti-V}
        \end{figure*}

\end{document}